\documentclass[12pt,preprint]{aastex}
\usepackage{epsf}

\def\nbody{$n$-body}
\def\deg{\ifmmode {^\circ}\else {$^\circ$}\fi}
\def\degree{\ifmmode {^\circ}\else {$^\circ$}\fi}
\def\mum{\ifmmode {\rm \,\mu {\rm m}}\else $\rm \,\mu {\rm m}$\fi}
\def\arcsec{\ifmmode ^{\prime \prime}\else $^{\prime \prime}$\fi}

\def\inch{\ifmmode ^{\prime \prime}\else $^{\prime \prime}$\fi}
\def\gs{\ifmmode {{\rm g~s^{-1}}}\else ${\rm g~s^{-1}}$\fi}
\def\msunyr{\ifmmode {M_{\odot}~{\rm yr^{-1}}}\else $M_{\odot}~{\rm yr^{-1}}$\fi}
\def\msun{\ifmmode {M_{\odot}}\else $M_{\odot}$\fi}
\def\rsun{\ifmmode {R_{\odot}}\else $R_{\odot}$\fi}
\def\lsun{\ifmmode {L_{\odot}}\else $L_{\odot}$\fi}
\def\mstar{\ifmmode {M_{\star}}\else $M_{\star}$\fi}
\def\rstar{\ifmmode {R_{\star}}\else $R_{\star}$\fi}
\def\tstar{\ifmmode {T_{\star}}\else $T_{\star}$\fi}
\def\lstar{\ifmmode {L_{\star}}\else $L_{\star}$\fi}
\def\mwd{\ifmmode {M_{wd}}\else $M_{wd}$\fi}
\def\rwd{\ifmmode {R_{wd}}\else $R_{wd}$\fi}
\def\twd{\ifmmode {T_{wd}}\else $T_{wd}$\fi}
\def\lwd{\ifmmode {L_{wd}}\else $L_{wd}$\fi}
\def\md{\ifmmode {M_d}\else $M_d$\fi}
\def\ld{\ifmmode {L_d}\else $L_d$\fi}
\def\ad{\ifmmode A_d\else $A_d$\fi}
\def\ldlwd{\ifmmode L_d / L_{wd}\else $L_d / L_{wd}$\fi}
\def\ldlstar{\ifmmode L_d / L_\star\else $L_d / L_{\star}$\fi}
\def\rearth{\ifmmode {\rm R_{\oplus}}\else $\rm R_{\oplus}$\fi}
\def\mearth{\ifmmode {\rm M_{\oplus}}\else $\rm M_{\oplus}$\fi}
\def\qdstar{\ifmmode Q_D^\star\else $Q_D^\star$\fi}
\def\vsqd{\ifmmode v^2 / Q_D^\star\else $v^2 / Q_D^\star$\fi}
\def\kms{\ifmmode {\rm km~s^{-1}}\else $\rm km~s^{-1}$\fi}
\def\ms{\ifmmode {\rm m~s^{-1}}\else $\rm m~s^{-1}$\fi}
\def\vrel{\ifmmode v_{rel}\else $v_{rel}$\fi}
\def\mdot{\ifmmode \dot{M}\else $\dot{M}$\fi}
\def\mdotz{\ifmmode \dot{M}_0\else $\dot{M}_0$\fi}
\def\mesc{\ifmmode m_{esc}\else $m_{esc}$\fi}
\def\rmin{\ifmmode r_{min}\else $r_{min}$\fi}
\def\rmax{\ifmmode r_{max}\else $r_{max}$\fi}
\def\xmax{\ifmmode x_{max}\else $x_{max}$\fi}
\def\mmin{\ifmmode m_{min}\else $m_{min}$\fi}
\def\mmax{\ifmmode m_{max}\else $m_{max}$\fi}
\def\rmind{\ifmmode r_{min,d}\else $r_{min,d}$\fi}
\def\rmaxd{\ifmmode r_{max,d}\else $r_{max,d}$\fi}
\def\mmaxd{\ifmmode m_{max,d}\else $m_{max,d}$\fi}
\def\vrad{\ifmmode v_{rad}\else $v_{rad}$\fi}
\def\qz{\ifmmode q_{0}\else $q_{0}$\fi}
\def\qi{\ifmmode q_{i}\else $q_{i}$\fi}
\def\ql{\ifmmode q_{l}\else $q_{l}$\fi}
\def\qs{\ifmmode q_{s}\else $q_{s}$\fi}
\def\vhill{\ifmmode v_H\else $r_H$\fi}
\def\rhill{\ifmmode r_H\else $r_H$\fi}
\def\Rhill{\ifmmode R_H\else $R_H$\fi}
\def\rbrk{\ifmmode r_{brk}\else $r_{brk}$\fi}
\def\rdamp{\ifmmode r_{damp}\else $r_{damp}$\fi}
\def\rin{\ifmmode r_{in}\else $r_{in}$\fi}
\def\rout{\ifmmode r_{out}\else $r_{out}$\fi}
\def\tin{\ifmmode t_{in}\else $t_{in}$\fi}
\def\tout{\ifmmode t_{out}\else $t_{out}$\fi}
\def\ain{\ifmmode a_{in}\else $a_{in}$\fi}
\def\aout{\ifmmode a_{out}\else $a_{out}$\fi}
\def\r0{\ifmmode r_{0}\else $r_{0}$\fi}
\def\R0{\ifmmode R_{0}\else $R_{0}$\fi}
\def\m0{\ifmmode m_{0}\else $m_{0}$\fi}
\def\M0{\ifmmode M_{0}\else $M_{0}$\fi}
\def\xm{\ifmmode x_{m}\else $x_{m}$\fi}
\def\sigz{\ifmmode \Sigma_0\else $\Sigma_0$\fi}
\def\ergg{\ifmmode {\rm erg~g^{-1}}\else ${\rm erg~g^{-1}}$\fi}
\def\gyr{\ifmmode {\rm g~yr^{-1}}\else ${\rm g~yr^{-1}}$\fi}
\def\cms{\ifmmode {\rm cm~s^{-1}}\else ${\rm cm~s^{-1}}$\fi}
\def\gcms{\ifmmode {\rm g~cm^{-2}}\else $\rm g~cm^{-2}$\fi}
\def\gcmc{\ifmmode {\rm g~cm^{-3}}\else $\rm g~cm^{-3}$\fi}
\def\atil{\ifmmode {\tilde{a}}\else $\tilde{a}$\fi}
\def\ttil{\ifmmode {\tilde{t}}\else $\tilde{t}$\fi}
\def\sqrttt{\ifmmode {\tilde{t}^{1/2}}\else $\tilde{t}^{1/2}$\fi}

\def\orch{{\it Orchestra}}

\begin{document}

\title{Numerical Simulations of Collisional Cascades at
the Roche Limits of White Dwarf Stars}
\vskip 7ex
\author{Scott J. Kenyon}
\affil{Smithsonian Astrophysical Observatory,
60 Garden Street, Cambridge, MA 02138} 
\email{e-mail: skenyon@cfa.harvard.edu}

\author{Benjamin C. Bromley}
\affil{Department of Physics, University of Utah, 
201 JFB, Salt Lake City, UT 84112} 
\email{e-mail: bromley@physics.utah.edu}
%
%

\begin{abstract}

We consider the long-term collisional and dynamical evolution of solid 
material orbiting in a narrow annulus near the Roche limit of a white dwarf. 
With orbital velocities of 300~\kms, systems of solids with initial 
eccentricity $e \gtrsim 10^{-3}$ generate a collisional cascade where 
objects with radii $r \lesssim$ 100--300~km are ground to dust.  This 
process converts 1--100~km asteroids into 1~\mum\ particles in $10^2 - 10^6$~yr.  
Throughout this evolution, the swarm maintains an initially large vertical 
scale height $H$.  Adding solids at a rate $\dot{M}$ enables the system to 
find an equilibrium where the mass in solids is roughly constant. This 
equilibrium depends on $\dot{M}$ and \r0, the radius of the largest solid 
added to the swarm.  When \r0\ $\lesssim$ 10~km, this equilibrium is stable. 
For larger \r0, the mass oscillates between high and low states; the fraction 
of time spent in high states ranges from 100\% for large $\dot{M}$ to much 
less than 1\% for small $\dot{M}$.  During high states, the stellar luminosity 
reprocessed by the solids is comparable to the excess infrared emission 
observed in many metallic line white dwarfs.  

\end{abstract}


\section{INTRODUCTION}
\label{sec: intro}

Among nearby white dwarfs with H-rich (DA stars) or He-rich (DB stars) 
atmospheres, roughly 25\% have metallic absorption lines from O, Mg, Al, Si, 
Ca, and Fe \citep[e.g.,][and references therein]{zuckerman1998,zuckerman2010,
koester2014,kepler2015,kepler2016,farihi2016}. 
Stars in the DZ class have strong metallic lines with little or no H or He 
\citep[e.g.,][]{sion1990,koester2011,sion2014,kepler2015,kepler2016}.
A few per cent of these stars have near-IR excess emission from warm dust
which reprocesses roughly 1\% of the stellar luminosity 
\citep[e.g.,][]{kilic2005,reach2005,hansen2006,kilic2006,tremblay2007,
vonhippel2007, kilic2008,farihi2009,girven2011,debes2011,chu2011,girven2012,
barber2012,hoard2013,bergfors2014,rocchetto2015,barber2016,bonsor2017}. 
A few systems also have metallic emission features which sometimes display 
the characteristic double-peaked profile of a circumstellar disk 
\citep[e.g.,][]{gansicke2006,gansicke2007,gansicke2008,melis2010b,
farihi2012,melis2012a,debes2012b,wilson2014}.

Detailed models demonstrate that heavy metals in white dwarf atmospheres
are continually replenished. Time scales for metals to diffuse from 
the atmosphere into the core are much shorter than the time scale 
for the white dwarf to cool \citep[e.g.,][]{fontaine1979,alcock1980a,
lacombe1983,wesemael1984,dupuis1992,althaus2000,koester2009}. 
Explaining the observed abundances requires time-averaged accretion 
rates of $10^5 - 10^{12}$~\gs\ for white dwarfs with ages of 0.1--3~Gyr
\citep[e.g.,][]{koester2006,deal2013,koester2014,farihi2016}. To put
these rates in perspective, steady accretion at $10^{10}$~\gs\ adds
a km-sized object to the white dwarf every 2--3 weeks. Sustaining this
rate for 1~Gyr requires a reservoir of 0.05~\mearth\ \citep[see also 
Figs. 10--11 of][]{farihi2016}.

There are two sources for accreted material: the interstellar medium
and solids leftover from a planetary system \citep[e.g.,][]{lacombe1983,
aanne1985,alcock1986,aanne1993,dupuis1993,jura2003,koester2006,jura2007a,
jura2007b,wyatt2014}. As summarized in \citet{farihi2016}, various 
observations rule out accretion from the ISM. In the current paradigm, 
solids on roughly circular orbits at large $a$ survive the evolution of
the central star into a red giant, ejection of a planetary nebula, and
contraction into a white dwarf \citep[e.g.,][]{stern1990,parriott1998,
debes2002,villaver2007,dong2010,bonsor2011,veras2013,mustill2014}.  
These solids are then somehow perturbed onto very high eccentricity 
($e \gtrsim 0.99$) orbits which pass within the Roche limit of the 
white dwarf \citep[e.g.,][]{jura2003,jura2008,debes2012a,veras2013}. 
Although the solids might hit the white dwarf directly, tidal forces 
probably disrupt the solids into myriad pieces which collide, fragment,
and vaporize \citep[see also][and references therein]{brown2017}. 
Over time, various physical processes somehow place material on nearly 
circular orbits close to the Roche limit, which then accretes onto 
the white dwarf.

Despite numerous investigations into the delivery of solids close 
to the white dwarf \citep[e.g.,][]{debes2002,veras2015a,bonsor2015,
antoniadou2016,payne2017,brown2017,petrovich2017} and the structure 
and evolution of solids and gas on circular orbits
\citep[e.g.,][]{rafikov2011a,rafikov2011b,bochkarev2011,metzger2012,
rafikov2012}, few studies focus on the physical processes which 
convert very high $e$ orbits into nearly circular orbits 
\citep[see also][]{veras2014a,veras2015b}. 
Following tidal disruption, various processes -- including collisions 
and gravitational stirring among the solids, vaporization, and 
interactions between the solids, the gas, and the stellar radiation 
field -- change the orbits and physical properties of the solids. 
Currently, there is no single calculation which follows all of these 
processes in detail.

In this paper, we begin to consider how solid material on very high $e$ 
orbits passing very close to a central white dwarf makes its way into 
the white dwarf photosphere. Because outcomes of various delivery models 
are uncertain, we study an idealized model where solids reside in a 
narrow annulus orbiting with initial eccentricity $e_0$ at the Roche 
limit. Our goal is to learn whether collisional and dynamical processes 
convert a system with non-zero $e_0$ into one with $e$ very close to zero.

For this initial study, we perform a suite of numerical simulations
which include {\it only} collisional and dynamical processes within 
the solids, Poynting-Robertson (PR) drag, and tidal interactions with 
the central star. Although these calculations ignore gas dynamics, 
simple estimates suggest gas drag probably has modest impact on the 
evolution.  With this focus, we constrain the impact of collisional 
damping, dynamical friction, PR drag, and viscous stirring in setting 
the orbital properties of solids within the tidal field of the white 
dwarf.

Aside from establishing the long-term evolution of solids at the Roche 
limit, these calculations make initial predictions for the magnitude 
and behavior of IR excess emission as a function of the accretion rate 
and other properties of the solids. The second goal of this study is 
to begin to understand whether some aspects of our simulations can
explain trends in the observations of metallic line white dwarfs.

Our investigation begins with some theoretical background to motivate 
the initial conditions for a suite of numerical calculations (\S2).  
After describing our algorithms and summarizing our results (\S3), 
we discuss the likely impact of gas dynamics and radiative processes,
place our approach in context with other theoretical studies, make 
preliminary connections to observations, and outline future steps (\S4).  
We conclude with a brief summary (\S5).

\section{THEORETICAL BACKGROUND}
\label{sec: back}

In the standard theoretical picture, solid fragments orbiting within the 
Roche limit dynamically relax into a flat disk \citep[e.g.,][and references 
therein]{debes2002,farihi2016}. To avoid rapid depletion by PR drag, the 
disk is radially extended, vertically thin, and optically thick 
\citep[e.g.,][]{jura2003,rafikov2011a,farihi2016}.  Adopting a radial 
temperature distribution for a flat disk, $T \propto r^{-3/4}$ 
\citep[e.g.,][]{fried1985,adams1987,kh1987,chiang1997}, enables fits of 
model disk fluxes to observations of many systems with IR excesses 
\citep[e.g.,][and references therein]{jura2003,farihi2016}.  The small 
vertical extent of the disk, $H \approx$ 1~cm to 1~m, implies a tiny 
vertical velocity dispersion, $v \lesssim$ 0.1~\cms; if $i/e$ = 0.5,
the orbital eccentricity, $e \lesssim 10^{-8}$. We want to learn whether 
ensembles of particles on eccentric orbits with $e \approx 10^{-3}$--1 
can reach the low $e$ required by this model.

As a reasonable starting point for this discussion, we examine the 
evolution of solids orbiting near the Roche limit with initial 
eccentricity $e_0$ and inclination $\imath_0$. The solids have initial 
mass $M_0$ and an initial size distribution between a minimum radius 
\rmin\ and a maximum radius \r0.  As the system evolves, additional 
solids are input at a rate \mdotz.  These solids have radius \r0\ and 
orbital parameters $e_0$ and $\imath_0$.  Throughout the calculation, 
\rmin\ is held fixed; the radius of the largest object in the grid, 
\rmax, changes as collisions add or remove mass.  Our goal is to 
establish outcomes of collisional evolution as a function of $e_0$, 
\r0, and \mdotz.

To set $e_0$ and $\imath_0$, we derive an orbital eccentricity $e_V$ where 
high velocity collisions between solid particles vaporize a negligible 
amount of material. Unless gravitational interactions among the solids 
raise $e$ and $\imath$, starting with $e_0 \le e_V$ and $\imath_0 = e_0/2$ 
ensures that collisions produce little or no gas throughout the evolution.  
The orbital velocity is $v_K = \sqrt{G \mwd / a} $ $\approx$ 300--350~\kms\ for 
white dwarf mass \mwd\ = 0.6~\msun\ and semimajor axis $a \approx$ 1~\rsun. 
Simulations suggest collisions with impact velocities 
$\lesssim$ 3~\kms\ ($\gtrsim$ 30~\kms) yield debris with a gas 
content $\lesssim$ 1\% ($\gtrsim$ 20\%--30\%) of the initial mass 
\citep{tielens1994,mann2005,czech2007}. Thus, we adopt $e_0$ = 
$e_V \approx$ 0.01. 

To set \r0, we rely on data for asteroids in the solar system, which have 
radii ranging from 100--300~km to $\lesssim$ 0.5--1~km 
\citep[e.g.,][]{yoshida2003,yoshida2007,gladman2009}. To sample this range, 
we consider \r0\ = 0.1--100~km. As outlined below, sublimation probably 
sets a lower limit on the size of solid particles orbiting near the white 
dwarf, \rmin\ = 0.1--1~\mum.  Coagulation calculations are rarely sensitive 
to \rmin, so we set \rmin\ = 1~\mum\ and do not consider other values.
Our choices for \mdotz, $10^{7} - 10^{13}$ \gs, are based on accretion 
rates inferred from the abundances of metals in white dwarf atmospheres, 
$10^{5} - 10^{12}$ \gs\ \citep[][and references therein]{farihi2016}.

Although we expect \r0\ and \mdotz\ to vary sporadically in a real system, 
these variables are held constant in each simulation. Setting \mdotz\ = 0 
allows us to investigate systems where the mass flow rate through the 
cascade is much faster than the input rate. In calculations with 
\mdotz\ $\gg$ 0, our goal is to learn whether the cascade finds a 
steady-state. Simulations over the complete range of \mdotz\ allow 
us to constrain time scales for evolving from one state to another and
the detectability of cascades as a function of \r0\ and \mdotz. The
range we consider is sufficient to extrapolate results to other choices 
and to infer the impact of a time-varying \r0\ or \mdotz.

\subsection{Coagulation Code}
\label{sec: back-code}

To follow the evolution of rocky solid particles orbiting a white dwarf,
we rely on \orch, a parallel \verb!C++/MPI! hybrid coagulation + 
\nbody\ code that tracks the accretion, fragmentation, and orbital 
evolution of solid particles ranging in size from a few microns to 
thousands of km \citep{kenyon2002,kb2008,bk2011a,kb2016a,knb2016}. 
The ensemble of codes within \orch\ includes a multi-annulus coagulation 
code for small particles, an \nbody\ code for large particles, and a 
radial diffusion code to follow the evolution of a gaseous circumstellar 
disk.  Other algorithms link the codes together, enabling each component 
to react to the evolution of other components.

In this study, we assume particles lie within a single annulus of width 
$\Delta a$ at a distance $a$ from the central star ($\Delta a = 0.2 a$). 
Within the annulus, there are $M$ mass batches with characteristic mass $m_i$ 
and logarithmic spacing $\delta = m_{i+1} / m_i$; adopting $\delta$ = 1.4 
provides a reasonably accurate solution for the cascade 
\citep[e.g,][and references therein]{kb2015a,kb2015b,kb2016a}. 
Batches contain $N_i$ particles with total mass $M_i$, average mass 
$\bar{m}_i = M_i / N_i$, horizontal velocity 
$h_i$ ($e_i$ = $\sqrt{1.6} h_i / v_K$), and vertical velocity 
$v_i$ (sin~$\imath$ = $\sqrt{2} v_i / v_K$).  The number of particles, 
total mass, and orbital velocity of each batch evolve through physical 
collisions and gravitational interactions with all other mass batches in 
the ring. 

To specify collision rates, we adopt the particle-in-a-box algorithm. In 
this approach, the collision rate of all particles $i$ with particles $j$
is $\dot{N}_i$ = $N_i ~ N_j ~ \sigma ~ v ~ f_g ~ \epsilon / V$, where
$\sigma$ is the geometric cross-section, $v$ is the relative velocity,
$f_g$ is the gravitational focusing factor, $V$ is the volume occupied 
by the particles, and $\epsilon$ is a factor to avoid double counting when
$i = j$ \citep{kl1998,kb2001,kb2002a}.  The relative velocity depends on 
$h_i$ and $v_i$.  When relative velocities are large (small), $f_g$
is derived in the dispersion (shear) regime with tidal effects included 
\citep{kl1998,kb2004a,kb2012}.

Collision outcomes depend on the ratio of the center-of-mass collision
energy $Q_c$ to the collision energy required to eject half of the mass
to infinity \qdstar. When two particles collide, the instantaneous mass 
of the merged particle is
\begin{equation}
m = m_i + m_j - \mesc ~ ,
\label{eq: msum}
\end{equation}
where the mass of debris ejected in a collision is
\begin{equation}
\mesc = 0.5 ~ (m_i + m_j) \left ( \frac{Q_c}{Q_D^*} \right)^{b_d} ~ ,
\label{eq: mej2}
\end{equation}
the center-of-mass collision energy is
\begin{equation}
Q_c = { m_i ~ m_j ~ v^2 \over 2 ~ (m_i + m_j)^2 } ~ ,
\label{eq: qc}
\end{equation}
the binding energy of a merged pair of particles is
\citep{benz1999,lein2012}
\begin{equation}
\qdstar = Q_b r^{\beta_b} + Q_g \rho_p r^{\beta_g} ~ ,
\label{eq: qdstar}
\end{equation}
and $b_d$ is a constant of order unity. In the expression for \qdstar,
the first (second) term corresponds to the bulk strength (gravity) 
component of the binding energy. We adopt $b_d$ = 1 and set 
fragmentation parameters in the \qdstar\ relation to those
appropriate for rocky material:
$Q_b \approx 3 \times 10^7$~erg~g$^{-1}$~cm$^{-\beta_b}$,
$\beta_b \approx -0.40$, $Q_g \approx$ 0.3~erg~g$^{-2}$~cm$^{3-\beta_g}$,
and $\beta_g \approx$ 1.35 for particles with mass density $\rho_p$
= 3~\gcmc\ \citep[see also][]{davis1985,hols1994,love1996,benz1999,
housen1999,ryan1999,arakawa2002,giblin2004,burchell2005}.  Particles 
in the debris have a power-law differential size distribution,
$N(r) \propto r^{-3.5}$. The mass of the largest particle 
in the debris is
\begin{equation}
\mmaxd = m_{L,0} ~ \left ( \frac{Q_c}{Q_D^*} \right)^{-b_L} ~ \mesc ~ ,
\label{eq: mlarge}
\end{equation}
$m_{L,0} \approx$ 0.01--0.5, and $b_L \approx$ 0--1.25 \citep{dohn1969,
weth1993,will1994,obrien2003,kb2008,koba2010a,weid2010,kb2016a}.
We adopt $m_{L,0} =$ 0.2 and $b_L \approx$ 1.

Near the Roche limit, the ability of colliding particles to merge into 
a larger particle depends on the tidal field of the white dwarf 
\citep[e.g.,][]{weiden1984a,ohtsuki1993,canup1995,karj2007,porco2007,
tisc2013,hyodo2014,yasui2014b}. We define the Hill radius 
\begin{equation}
\rhill = 
\left ( { m_i + m_j \over \mwd } \right )^{1/3} a ~ ,
\label{eq: rhill}
\end{equation}
which separates the volume where material is bound to the particles,
$r \lesssim \rhill$, from the volume where material is bound to the 
central star, $r \gtrsim \rhill$.
When the radius of the merged particle $r_s = r_i + r_j$ is smaller than
\rhill, the particles merge into a larger object. Otherwise, the collision
results in two particles with total mass $m$ from eq.~(\ref{eq: msum}). For
convenience, we scale the mass lost from each particle by the initial mass,
$m_{i,new}$ = $m_i (1 - \mesc / (m_i + m_j))$.

The orbital elements $e_i$ and $\imath_i$ evolve due to collisional damping 
from inelastic collisions, gravitational interactions, and interactions with 
the radiation field of the central star.  For inelastic and elastic collisions, 
we follow the statistical, Fokker-Planck approaches of \citet{oht1992} and 
\citet{oht2002}, which treat pairwise interactions (e.g., dynamical friction 
and viscous stirring) between all objects \citep[see also][and references
therein]{kl1998,kb2001,kb2002a,kb2004a,kb2004c,kb2008,kb2015a}. For short-range
interactions within 5--10 Hill radii, the Fokker-Planck formalism matches 
results from detailed $n$-body simulations. To treat interactions between 
solids with larger separations, we also calculate long-range stirring 
\citep{weiden1989}.  When the central star is a low luminosity white dwarf 
($\lwd \lesssim 10^{-2} ~ \lsun$), radiation pressure has negligible impact 
on the orbital elements.  However, we include radial drift and eccentricity 
damping from Poynting-Robertson drag \citep{burns1979}.  Our approach 
includes tidal terms appropriate for material within the Roche limit. 
Several test calculations with our algorithms reproduce results from 
previous studies of velocity evolution near the Roche limit 
\citep[e.g.,][]{canup1995,ohtsuki2000}.

Throughout the evolution, collisional disruption generates particles with 
radii smaller than \rmin, the smallest size included in the grid. We assume 
that these particles are removed by vaporization and do not interact with 
larger particles in the grid. Vaporized solids add material to a gaseous 
disk; over time, the gas accretes onto the white dwarf. We consider the 
likely impact of a gaseous on the solids in \S\ref{sec: disc-drag}.

\subsection{Tidal Disruption}
\label{sec: back-tides}

Aside from treating tidal effects in collision outcomes and dynamical evolution,
we must consider tidal disruption of solids near the Roche limit. Among various
options for analyzing tidal stability \citep[e.g.,][]{aggarwal1974,dobrovolskis1990,
davidsson1999,davidsson2001}, we infer constraints using failure criteria developed
for terrestrial soils and applied to satellites of Mars and the giant planets in the
solar system \citep[e.g.,][]{holsapple2006,holsapple2008,sharma2009,sharma2014}. 
As in \citet{holsapple2006,holsapple2008}, we divide solids into small objects
with finite cohesiveness and negligible self-gravity and large objects dominated by 
gravity. Approaches outlined in \citet{sharma2009,sharma2014} yield similar results.

For small objects, the upper limit of the spin periods of asteroids as a function 
of their measured diameter implies a cohesiveness 
$k = 2.25 \times 10^7 r^{-1/2}$ dyne~cm$^{-2}$. Larger objects have smaller 
cohesiveness (more faults in their structure). Applying the Drucker-Prager model
outlined in \citet{holsapple2008}, we calculate $(a_d/\rwd) (\rho / \rho_{wd})^{1/3}$, 
where $a_d$ is the minimum stable distance for a solid with a given shape, spin,
and mean density $\rho$ orbiting a more massive central object with radius
\rwd\ and mean density $\rho_{wd}$.  For simplicity, we assume a spin axis 
parallel to the orbital axis.

\begin{figure} 
\includegraphics[width=6.5in]{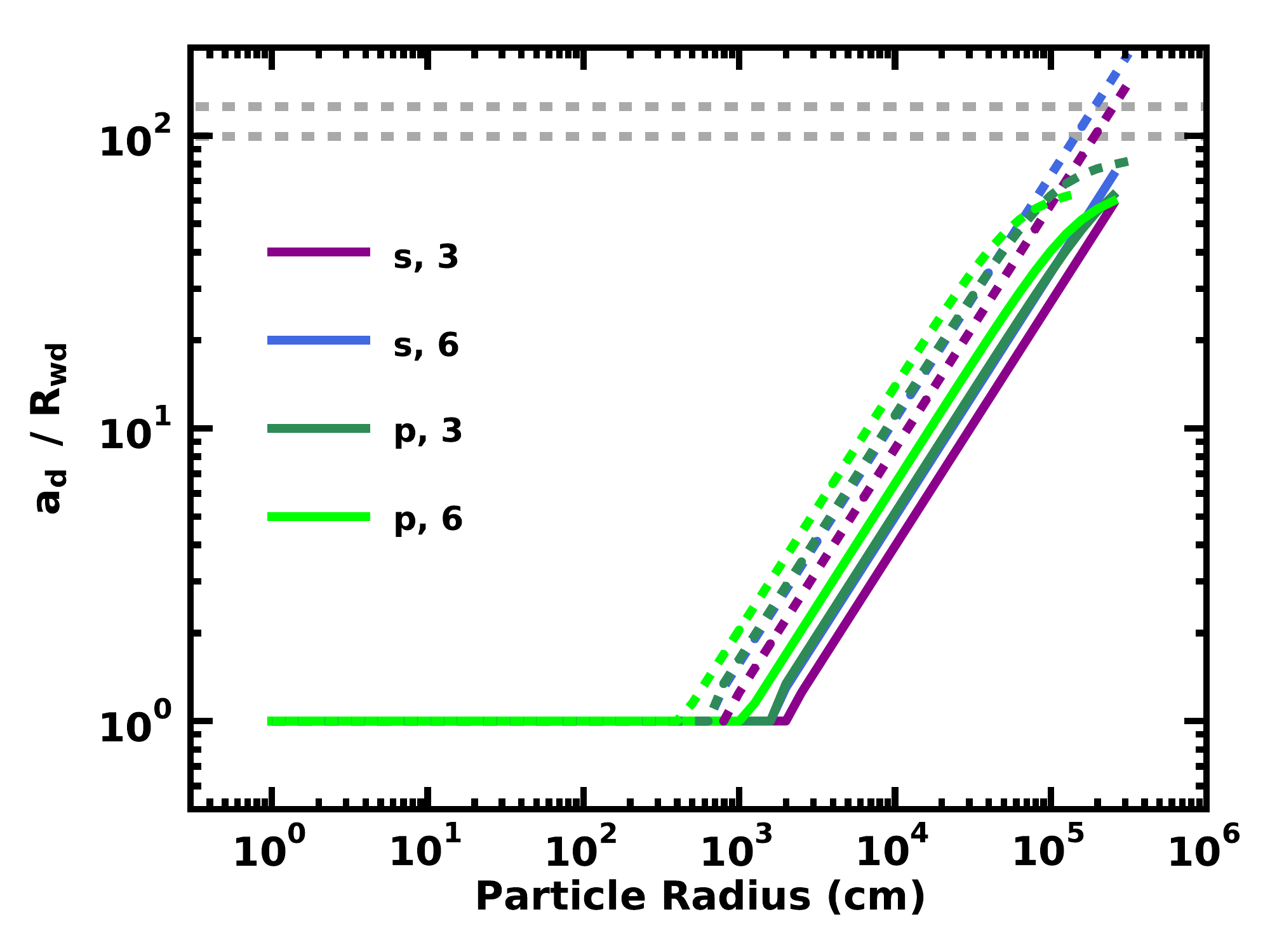}
\vskip 3ex
\caption{
Innermost stable orbits for solid particles with cohesive strength
orbiting a white dwarf with a mass of 0.6~\msun\ and a radius of
1.4~\rearth. The solids have rotational periods equal to their
orbital periods.
Dashed horizontal grey lines indicate the Roche limit for fluids 
with $\rho$ = 3~\gcmc\ and 6~\gcmc.
As indicated in the legend, colored lines indicate results for
spherical (`s') or prolate (`p', 2:1:1) particles with mean density
$\rho$ = 3 or 6~\gcmc. 
Particles with radii smaller than 10~m (1~km) are stable against 
tidal disruption at the surface of the white dwarf 
(at $a \le 40~\rwd$).
\label{fig: roche1}
}
\end{figure}

Fig.~\ref{fig: roche1} shows results for solids with $r$ = 1~cm to 10~km.  
The dashed grey lines denote the classical Roche limit, $a_R / \rwd$, for 
fluids with $\rho$ = 3~\gcmc\ (upper curve) and $\rho$ = 6~\gcmc\ (lower 
curve) orbiting a 0.6~\msun\ white dwarf with a mean density\footnote{To derive
$\rho_{wd}$, we adopt the mass radius relation of \citet{verbunt1988}.  Other 
options yield similar results.} $\rho_{wd} = 4.45 \times 10^5$~\gcmc, where
\begin{equation}
{a_R \over \rwd } \approx C_R 
\left ( { \rho_{wd} \over 10^6~\gcmc\ } \right )^{1/3}
\left ( { 3~\gcmc\ \over \rho } \right )^{1/3} ~ 
\label{eq: roche}
\end{equation}
and $C_R \approx$ 170.  Adopting the standard cohesiveness (solid curves), 
solids with $r \le$ 3--10~km are stable inside the classical Roche limit.
Formally, particles with radii $r \le$ 10~m are stable even when they orbit 
at the surface of the white dwarf. Larger objects are progressively less 
stable until $r \approx$ 1--10~km, when the stability limit approaches the 
Roche limit. Prolate solids are more stable than spherical particles. 
Factor of ten smaller cohesiveness (dashed lines) has little impact on 
the results.  

Other approaches to deriving the Roche limit for small solids yield fairly 
similar results. In his landmark paper, \citet{jura2003} adopted an expression 
for the Roche limit from \citet{davidsson1999} and set\footnote{Here, we 
correct for a factor of 10 overestimate of $\rho_{wd}$ in Jura's (2003) 
analysis, which places $a_d$ for an asteroid outside $a_R$.} the disruption 
radius $a_d / \rwd$ as roughly 40\% of $a_R / \rwd$.  In more recent studies, 
$a_d \approx $ 80\% \citep{bear2013} to 120\% \citep{veras2014a} of 
$a_R$ \citep[see also][]{farihi2016}. All of these investigations assume the
solids are rigid spheres, which tends to overestimate the disruption radius
\citep[e.g.,][]{davidsson1999,davidsson2001,holsapple2008,sharma2014}.

In the gravity limit, 10--100~km objects are probably also stable at 
$a_d \approx$ 0.5--0.7 $a_R$. In the solar system, the small satellites 
of Mars and the gas giants are tidally stable at approximately 2/3 of 
the standard fluid limit.
To explain observations of eclipses in WD1145+017, \citet{veras2017} simulated
asteroid disruption with the $N$-body code PKDGRAV \citep[e.g.,][and references 
therein]{lein2000,richard2000,richard2005}. With $a \approx 0.7 a_R$, spherical
objects with $r \approx$ 1--100~km and $\rho \gtrsim$ 3~\gcmc\ are stable against 
tidal disruption on 3 month to 2 yr time scales. Less dense solids disrupt within
1--30~days. For a fixed density, more massive asteroids are more stable.  

Based on this discussion, we focus on the evolution of tidally stable objects 
with $r \le$ 100~km.  Formally, tidally stable solids orbiting the white dwarf 
are prolate ellipsoids with aspect ratios of roughly 2:1:1. Tidal distortion has 
no impact on stirring rates, which depend on particle masses.  Collision rates 
depend on geometric cross-sections, however, we ignore the slight changes in 
rates for ellipsoidal particles in this initial study.

\subsection{Strategy for Numerical Simulations}
\label{sec: back-strategy}

In the next section, we consider a suite of numerical simulations designed to
infer whether collisional evolution in systems of solids with $e_0$ = 0.01 and 
large vertical scale height leads to states with circular orbits and negligible
vertical scale height. First, we demonstrate in \S\ref{sec: sims-damp} that 
swarms of {\it indestructible mono-disperse particles} with the cross-sectional 
area required to explain the IR excess emission of white dwarf debris disks 
damp on very short time scales. 

We then consider a standard cascade calculation of a swarm of solids with 
initial mass $M_0$ and no input mass from external sources (\S\ref{sec: 
sims-casc1}).  Although these systems also evolve rapidly, they never reach a 
state with small $e$ and negligible $\imath$: collisional damping is negligible. 

This failure motivates the model described in \S\ref{sec: sims-casc2} where 
the input rate of solids is sufficient to balance the loss rate of particles 
with $r < \rmin$. When the input particles are small, $\rmax \lesssim$ 10--30~km, 
collisional evolution yields an equilibrium mass proportional to the mass input 
rate \mdotz. Swarms with \rmax\ $\gtrsim$ 10--30~km cycle through periods of 
high mass and near-zero mass.  The high mass limit of these swarms scales with 
\mdotz. Once again, though, the swarms never attain a state with small $e$ and 
negligible vertical scale height.

\section{NUMERICAL SIMULATIONS}
\label{sec: sims}

To guide our interpretation of the calculations, we set several useful parameters.
For an optically thin, vertically extended ring of debris, the luminosity 
reprocessed by small particles is 
\ldlwd\ = $A_d / 4 \pi a^2$ where \ad\ is the total cross-sectional area and $a$ 
is the distance of the ring from the central star. Setting $a_0 \approx 1~\rsun$ 
and $A_d \approx 10^{21}$~cm$^2$ yields \ldlwd\ = 0.022; observations indicate 
$\ldlwd \approx 10^{-3}$--0.03 \citep[e.g.,][and references therein]{barber2012,
hoard2013,bergfors2014,rocchetto2015,barber2016,farihi2016}.  If the debris consists 
of mono-disperse particles with radius $r$, the total mass is $\md = 4 \rho r A_d / 3$
= $1.33 \times 10^{21} \rho r$~g.

As derived in \citet{kb2016a}, the collision time for a system of mono-disperse 
particles orbiting within a ring is $t_0 = r \rho P / 12 \pi \Sigma$, where 
$\rho$ is the mass density, $P = 2 \pi / \Omega$ is the orbital period, and 
$\Sigma = M_d / 2 \pi a \Delta a$ is the surface density \citep[see also][and 
references therein]{knb2016}. This derivation assumes $f_g$ = 1.  For our 
standard ratio of \md\ to \ad:
\begin{equation}
t_0 \approx 1.6 \times 10^3~{\rm s}
\left ( { 10^{21} ~ {\rm cm^2} \over A_d } \right )
\left ( { \Delta a \over 0.2 a } \right ) 
\left ( { a \over 1~\rsun } \right )^{7/2}
\label{eq: t0}
\end{equation}
Near the Roche limit, the collision time for mono-disperse particles with a
detectable IR excess is roughly 10\% of the orbital period around a 
0.6~\msun\ white dwarf, $P = 1.3 \times 10^4$~s ($a$ / 1~\rsun)$^{3/2}$.

When collisions produce a broad size distribution of particles, the collision
time becomes smaller \citep{wyatt2008,koba2010a,wyatt2011,kb2016a}. Following
previous studies, we set the collision time as $t_c = \alpha t_0$. In cascades
where \qdstar\ is independent of particle radius, $v$ is the collision velocity,
and $v^2 / \qdstar \gg$ 1, $\alpha \approx 21 (v^2 / \qdstar)^{-0.8}$
\citep{kb2017}. White dwarf debris disks composed of 1~cm objects with 
$e \approx$ 0.01 have $v^2 / \qdstar\ \approx$ 100. The collision time
$t_c$ is then a factor of five smaller than the mono-disperse collision 
time $t_0$.

For many conditions, PR drag operates on time scales much longer than 
physical collisions or gravitational stirring. Adopting relations for
the time derivatives in semimajor axis, $\dot{a}$, and eccentricity, 
$\dot{e}$ \citep{burns1979}, the ratio $ (\dot{e} / e)~(a / \dot{a})$ 
is much smaller than 1 when $e \approx$ 1 and somewhat larger than 1 
when $e \approx$ 0. In our simulations with $e \approx$ 0.01, the time 
scale for radial drift is then roughly equal to the time scale for 
eccentricity damping:
\begin{equation}
t_{PR} \approx 5~{\rm yr}
\left ( { r \over 1~\mu{\rm m} } \right )
\left ( { a \over 1~\rsun } \right )^2
\left ( { 10^{-2} \lsun\ \over \lwd } \right ) ~ .
\label{eq: tpr}
\end{equation}

When the surface density of solids becomes small, PR drag acts faster
than collisions or gravitational stirring. For swarms of 1~\mum\ (1~cm)
particles, the $t_{PR} \approx t_0$ 
when $\ad \approx 10^{16}$~cm$^2$ ($\ad \approx 10^{12}$~cm$^2$)
and $\ldlwd \approx 10^{-8}$ ($\ldlwd \approx 10^{-12}$).  In our 
calculations, these conditions are rarely met; thus, PR drag has little 
influence on outcomes.

Once collisions produce debris, small grains may sublimate before interacting
with other particles. Near the Sun, the sublimation time scale for 
1~\mum\ crystalline olivine grains is roughly 10~s at 8~\rsun, $10^3$~s at 
10~\rsun, and $10^5$~s at 11~\rsun\ \citep{kimura2002,mann2004,mann2006}. 
Although amorphous olivine grains sublimate much more rapidly, pyroxene 
grains of any type sublimate much more slowly.  Scaling these results to 
conditions appropriate for a white dwarf with an effective temperature 
$T_{wd} = 10^4$~K, 1~\mum\ crystalline olivine grains have a sublimation 
time of roughly $10^5$~s at the Roche limit. The time scale drops to 
$10^3$~s when $a \approx 0.9 a_R$. For comparison, pyroxene grains survive
for $10^3 - 10^5$~s at 0.4--0.5 $a_R$.

Although we do not know the grain type for debris disks around white dwarfs,
these results suggest grains with $r \gtrsim$ 2--3~\mum\ are safe from rapid 
sublimation at 0.5--1.0~$a_R$. Much smaller grains with $r \lesssim$ 
0.3~\mum\ probably sublimate before they collide with another particle. In 
between these two size ranges, the grains are about as likely to sublimate 
as to collide with another grain. For simplicity, we assume that grains with 
$r \lesssim$ \rmin\ = 1~\mum\ sublimate as soon as they are formed.

\subsection{Collisional Damping}
\label{sec: sims-damp}

To demonstrate how a swarm of solids {\it might} damp, we consider 
material\footnote{These choices are motivated by WD 1145+017 
\citep{vanderburg2015}, where solids with a large vertical scale height 
eclipse a central white dwarf.} with \ad\ = $10^{21}$~cm$^2$ orbiting 
at $a = 0.71 a_R$ = 1.15~\rsun\ ($P$ = 4.5~hr) around a 0.6~\msun\ white 
dwarf.  In these test simulations, we follow a mono-disperse set of 
{\it indestructible} solids which evolve by collisional damping and 
viscous stirring.  

\begin{figure} 
\includegraphics[width=6.5in]{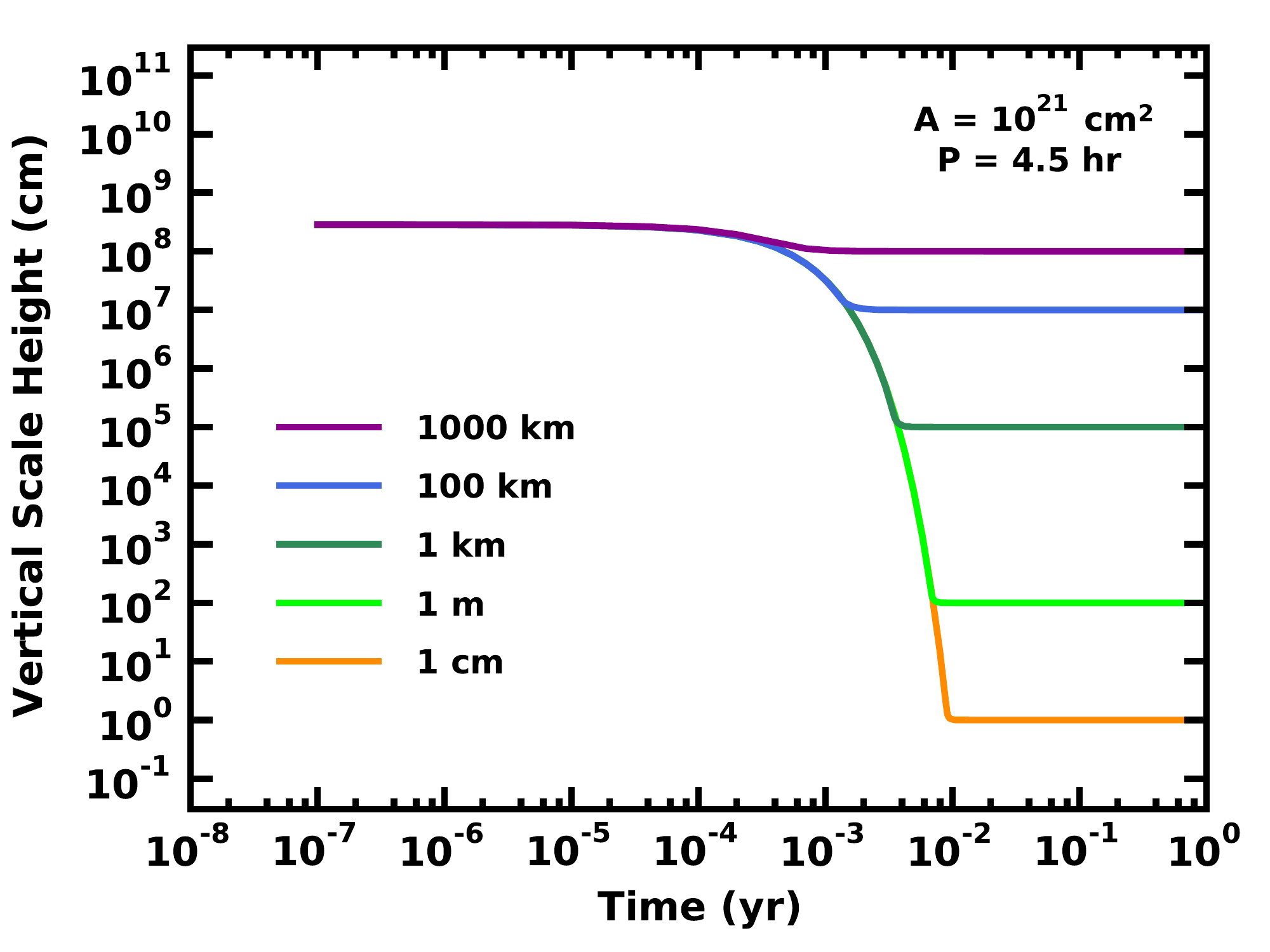}
\vskip 3ex
\caption{
Eccentricity decay from collisional damping for swarms of 
indestructible mono-disperse
solids with total cross-sectional area $A = 10^{21}$~cm$^2$ and orbital
period $P$ = 4.5 hr around a 0.6~\msun\ white dwarf. All swarms decay
on the same time scale to equilibrium eccentricities and vertical scale
heights that scale with the particle size, 
$e_{eq} \approx 1.7 \times 10^{-11}$ ($r$/1~cm) and $H \approx r$.
\label{fig: damp1}
}
\end{figure}

Fig.~\ref{fig: damp1} shows our results.  With no change in particle properties, 
the evolution proceeds until (i) collisional damping and viscous stirring 
balance or (ii) the vertical scale height is equal to 1--2 particle radii.  
Although damping times for swarms with identical \ad\ are identical, larger 
particles stir the swarm faster than smaller particles. Thus, swarms of 
1000~km particles find a balance with larger vertical scale height $H$ than 
swarms of 1~cm particles.  In this example, the equilibrium has 
$H \approx r$ and $e \approx 2 \times 10^{-11} r$.

As the vertical scale height of these rings approaches equilibrium, they 
are subject to gravitational instability \citep[e.g.,][and references 
therein]{gold1973,weiden1995,youdin2002,chiang2010}. For a mono-disperse 
set of particles orbiting a 0.6~\msun\ white dwarf, the critical scale 
height for instability is 
\begin{equation}
H_{crit} \approx 1~{\rm cm} 
\left ( { A_d \over 10^{21} ~ {\rm cm^2} } \right )
\left ( { r \over {\rm 1~cm}} \right )
\left ( { \rho \over {\rm 3~g~cm^{-3}}} \right )
\left ( { 0.2 a \over \Delta a } \right )
\left ( { 1~\rsun \over a } \right )^{1/2} ~ .
\end{equation}
When $H \lesssim r$ $a \approx$ 1~\rsun, systems with $A_d \approx$
$10^{21}$~cm$^2$ tend to be unstable. Closer to the white dwarf, 
swarms with somewhat smaller \ad\ are also unstable.

At the Roche limit of a white dwarf, massive swarms of indestructible solids 
find an equilibrium with a vertical scale height comparable to the particle 
radius. Lower mass swarms have larger $H$.  For reasonable total masses, the 
maximum $H$ is comparable to $r$ for $r \lesssim$ 1--10~cm and $\gtrsim$ 
5--10 $r$ for $r \gtrsim$ 10~cm.  Swarms of 1~cm or smaller particles with a 
total cross-sectional area large enough to produce an observable IR excess 
are probably gravitationally unstable. A system with enough mass in somewhat 
larger particles avoids the instability.  

Although these results are encouraging, solid particles on high $e$ orbits 
around a white dwarf are likely to suffer catastrophic collisions. We now 
consider whether systems with destructible particles can also find equilibria 
with small $H$.

\subsection{Collisional Cascades with No Mass Input}
\label{sec: sims-casc1}

\begin{figure} 
\includegraphics[width=6.5in]{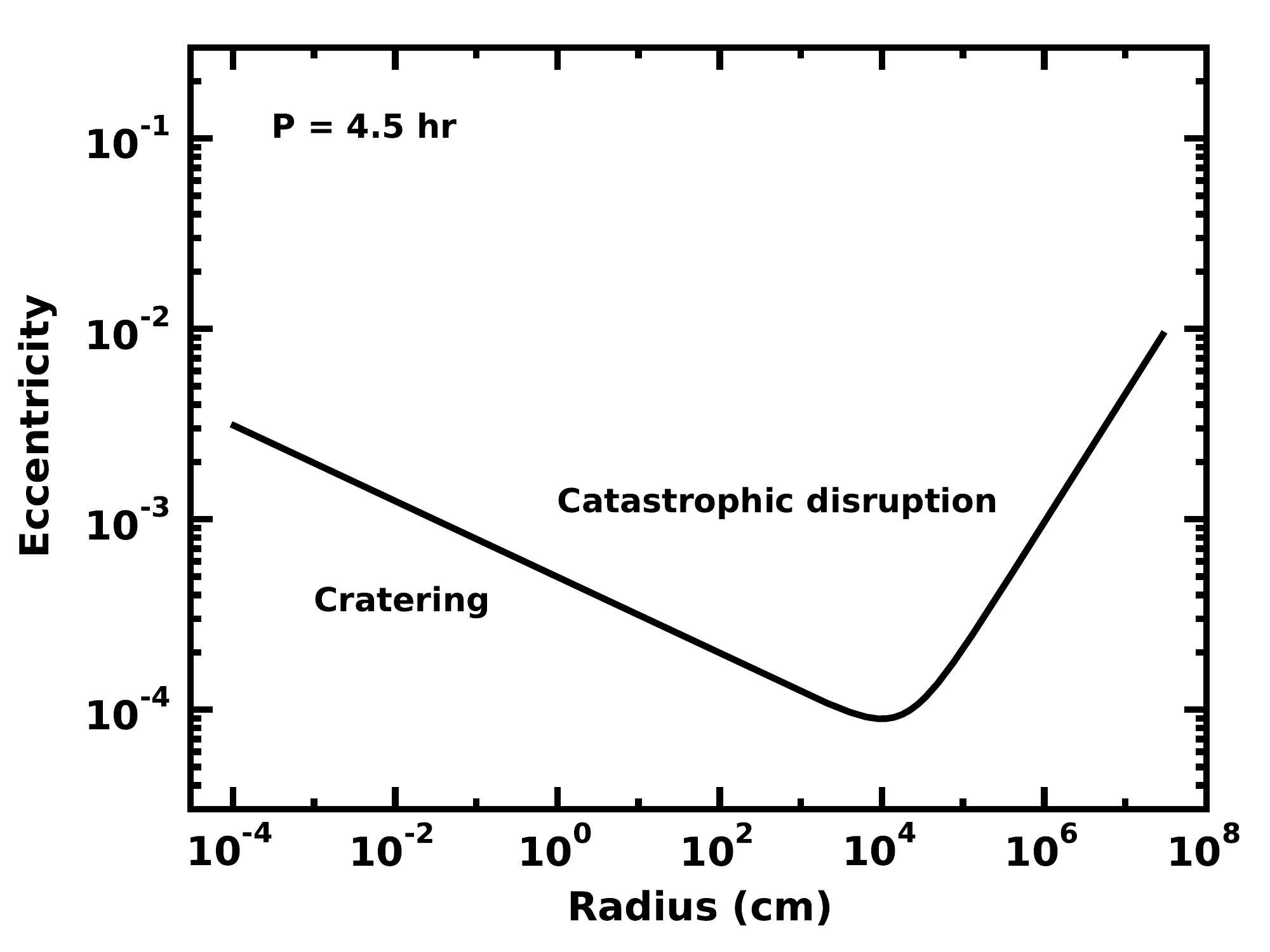}
\vskip 3ex
\caption{
Critical eccentricity, $e_c$ for catastrophic disruption. Collisions 
between equal mass objects with $e \ge e_c$ eject more than half of
the total mass in debris. When $e < e_c$, collisions eject less mass
and may produce a larger merged object.
\label{fig: qdstar}
}
\end{figure}

Particles orbiting near the Roche limit of a white dwarf are fairly 
easy to break.  We define an orbital eccentricity $e_c$ required for 
catastrophic disruption, where the collision ejects half the mass of 
the combined mass of two colliding particles. For two equal mass objects
with collision velocity $v \approx e v_K$, the center-of-mass collision 
energy is $Q_c = v^2/8 \approx e^2 v_k^2 / 8$. Setting $Q_c = \qdstar$ 
(with \qdstar\ from eq. \ref{eq: qdstar}) yields $e_c$.  For our 
\qdstar\ parameters, Fig.~\ref{fig: qdstar} shows $e_c$ for orbits with 
$P$ = 4.5~hr.  Starting with $e_0$ = 0.01 guarantees catastrophic 
disruption of all solids with radii between 1~\mum\ and 300~km.

To quantify the impact of destructive collisions on collisional damping,
we follow the evolution of swarms with initial surface density
$\Sigma_0$ = 100~\gcms\ and total mass $M_0 \approx 8 \times 10^{23}$~g
in an annulus centered at $a$ = 1.15~\rsun\ from a 0.6~\msun\ white dwarf.
Material in this annulus has orbital period $P$ = 4.5~hr, initial 
eccentricity $e_0$ = 0.01 and initial inclination $\imath_0 = e_0/2$. 
Calculations begin with a mono-disperse set of solids with initial radius
\r0.  

\begin{figure} 
\includegraphics[width=6.5in]{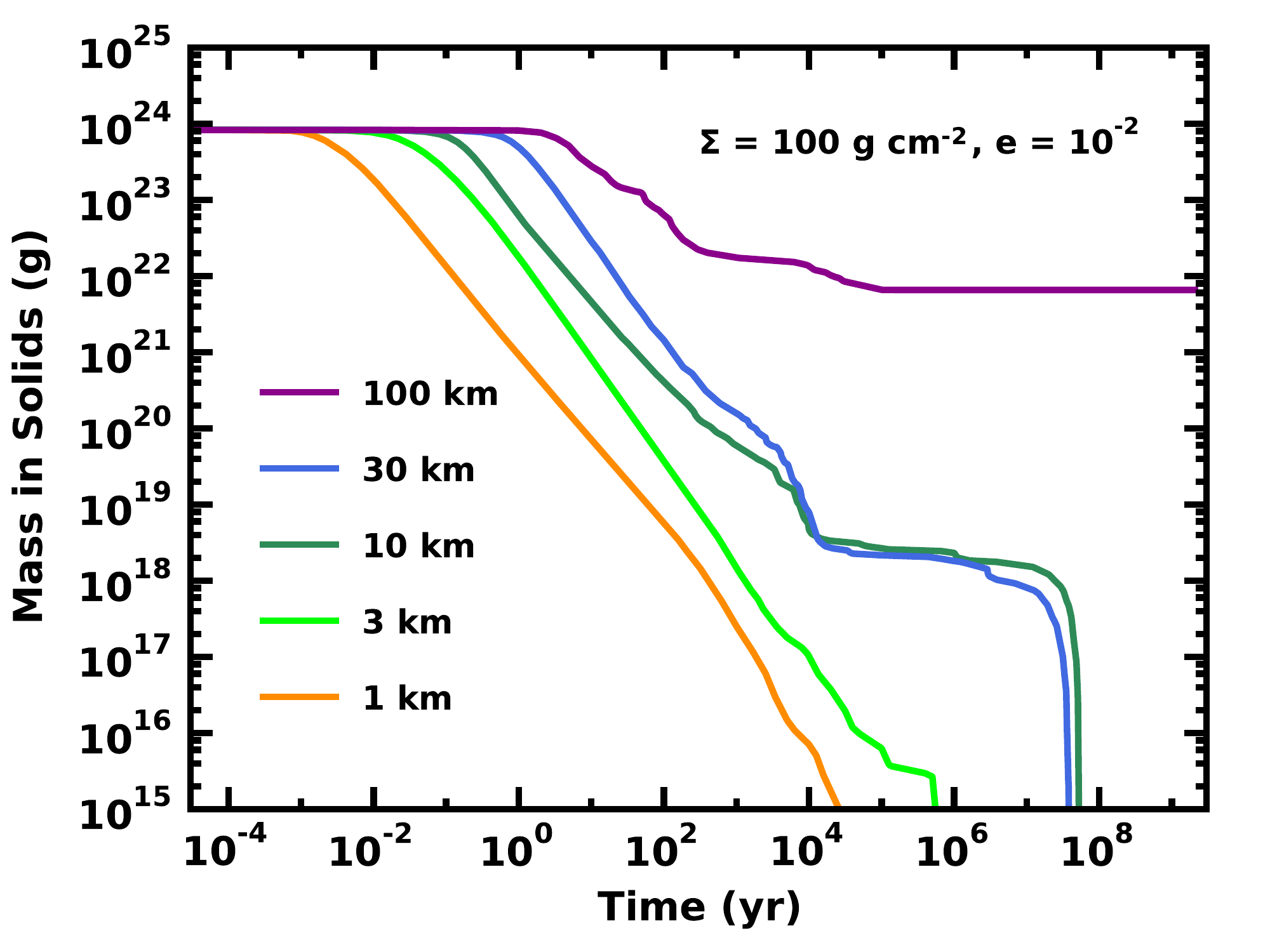}
\vskip 3ex
\caption{
Time evolution of the mass of initially mono-disperse swarms of solid 
particles orbiting a 0.6~\msun\ white dwarf with initial surface density 
$\Sigma_0$ = 100~\gcms\ and eccentricity $e$ = 0.01. The legend associates 
the solid curves with the initial radius \r0\ of the largest object.
As each system evolves, destructive collisions remove mass from the
system.  For \r0\ $\le$ 30~km, the mass drops rapidly to very low levels 
in $10^2 - 10^4$~yr. When \r0\ = 100~km, the cascade gradually reduces 
the swarm to a single 100~km object, which has no collisions after 
roughly 0.1~Myr.
\label{fig: mass1}
}
\end{figure}

Fig.~\ref{fig: mass1} illustrates the time evolution of the total mass 
in solids. In each calculation, it takes 1--2 collision times to generate 
copious amounts of small objects which systematically remove mass from 
larger objects. Because the collision time is proportional to the 
cross-sectional area of the swarm, systems composed of 1~km and smaller 
objects evolve faster than systems of 100~km and smaller objects.

Initially, the mass drops monotonically with time. As the calculation 
proceeds, all systems end up with 1 or 2 large objects which dominate 
the mass of the swarm. If the debris can slowly grind down these objects, 
the mass continues to drop with time. Sometimes, however, the debris 
evolves more rapidly than the largest solids. The mass then remains 
roughly constant in time until the remaining large particles collide 
with each other (if there are at least two of them, as in the tracks 
for calculations with \r0\ = 10~km and 30~km) or forever (if there is 
only one large object, as in the calculation with \r0\ = 100~km).

While the mass in each system declines, collisional damping among small 
particles with $r \approx$ 1~\mum\ to 10~cm overcomes stirring by the
largest solids.  However, the reduction in orbital $e$ and $\imath$ is 
rather small: $e$ ($\imath$) drops from 0.01 (0.005) to 0.004--0.006 
(0.002--0.003). At the end of these calculations, the vertical scale 
height $H \approx$ 1250--1500~km is roughly 10\% of the radius of the 
central white dwarf.

\begin{figure} 
\includegraphics[width=6.5in]{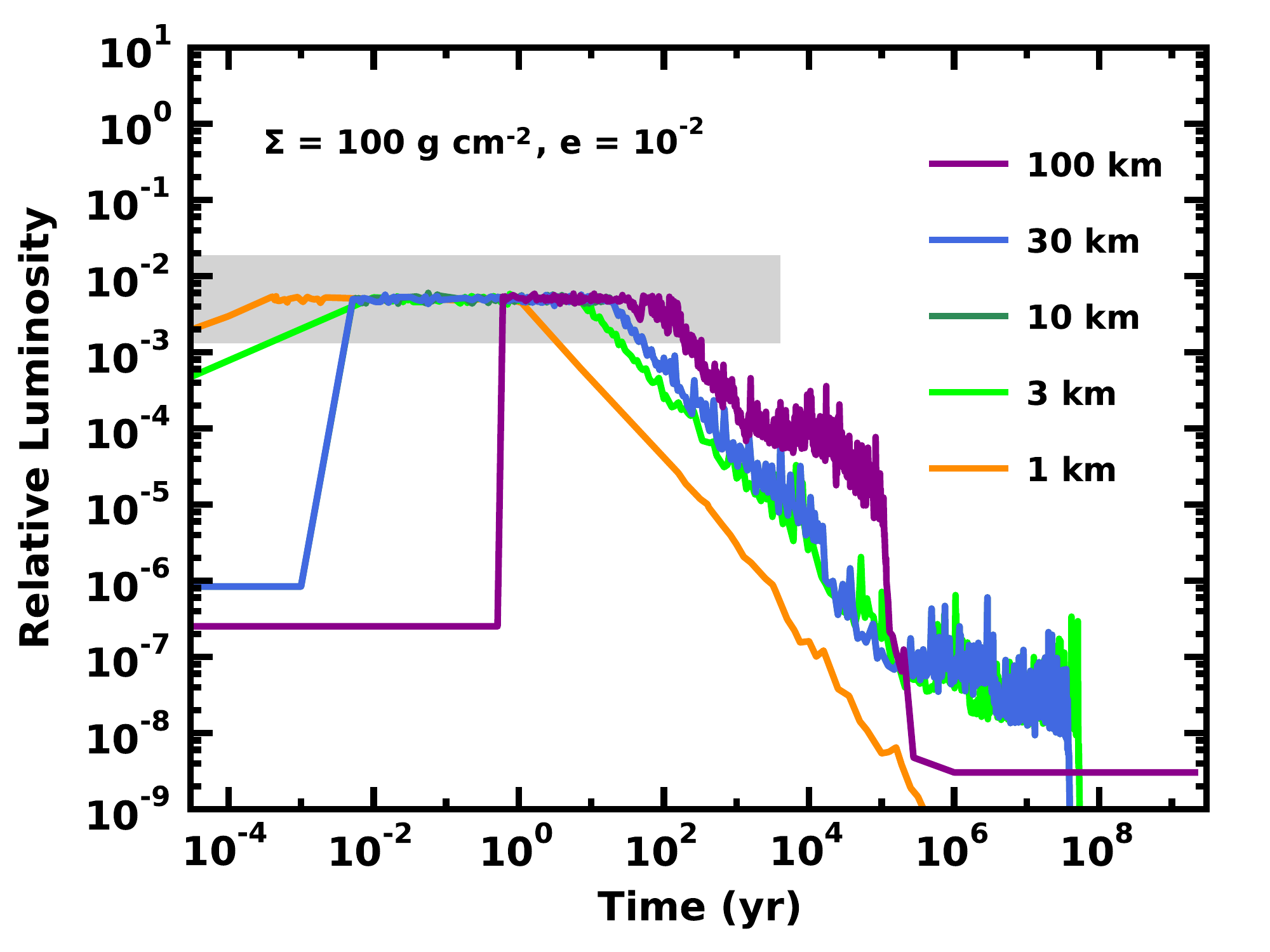}
\vskip 3ex
\caption{
As in Fig.~\ref{fig: mass1} for the reprocessed luminosity.  The horizontal
grey bar indicates the typical luminosity for the IR excesses of metallic 
line white dwarfs. After an initial spike of debris production, the luminosity 
drops to undetectable levels in $10^4$~yr or less. Systems where the initial 
\rmax\ is smaller decline more rapidly. Shot noise in collision rates produces
fluctuations in the luminosity.
\label{fig: lum1}
}
\end{figure}

For a short period of time, there is enough debris in any of these 
systems to produce an observable IR excess (Fig.~\ref{fig: lum1}). 
When \r0\ = 1--3~km, the IR luminosity maintains $\ldlwd\ = 0.01$ for 
1--10~yr. After 100~yr, \ldlwd\ falls well below observed levels. For 
larger \r0, the IR luminosity matches observed levels for 10--100~yr 
and then drops dramatically.

Occasional collisions among the largest objects produce the sporadic
spikes in \ldlwd. In systems with larger \r0, collisions among the
largest objects are less frequent and produce more debris. Thus, the
spikes in \ldlwd\ are more pronounced when \r0\ is larger.

Throughout the period when the IR luminosity is $10^{-2} - 10^{-3}$~\lwd,
the production rate of 1~\mum\ and smaller particles is $10^{12} - 10^{14}$~\gs.
This rate is somewhat larger than the inferred accretion rates of solids 
onto metallic line white dwarfs. If the 1~\mum\ and smaller particles 
sublimate, they will generate a gaseous ring which then expands into a 
disk.  The rate of accretion onto the white dwarf depends on the viscous 
time scale and the underlying structure of the accretion disk. Deriving 
the rate of accretion onto the central white dwarf requires the solution 
of the radial diffusion equation for the gas \citep[e.g.,][]{metzger2012} 
which is beyond the scope of the present effort.

These calculations demonstrate that collisional damping is ineffective in
reducing the vertical scale height of ensembles of solid particles with
\r0\ = 1--100~km and finite \qdstar. Tests with \r0\ = 0.1--0.3~km or 
\r0\ = 300--1000~km yield similar results. Factor of three changes in
\qdstar\ also have modest impact on the evolution. Adopting smaller
(larger) values for \qdstar\ slows down (speeds up) the decline in \md;
however, the overall character of the evolution is unchanged. In all cases, 
destructive collisions reduce the mass in solids to nearly zero before
collisional damping can reduce the vertical scale height dramatically.

In these systems, the production rate of small particles and the dust
luminosity are only briefly comparable with those required by observations.
During these short periods, however, the evolution time is much shorter
than typical time scales observed in metallic line white dwarfs.

\subsection{Collisional Cascades with Mass Input}
\label{sec: sims-casc2}

If the material lost to 1~\mum\ particles is continuously re-supplied by an
external source, it might be possible to maintain an equilibrium mass and 
luminosity for a ring of solid particles. In this equilibrium, the rate of 
mass input at the upper mass end of the cascade balances the rate the cascade 
generates solids with $r \le \rmin$.  Defining $\dot{M}$ as the mass loss rate 
through the cascade, $\dot{M} = M_d / t_c$.  Using $t_c = \alpha t_0$,
$M_{d, eq} = (\alpha \r0 \rho P a \Delta a \dot{M} / 6)^{1/2}$. To derive a 
simple closed form for the equilibrium mass in the gravity regime for 
\qdstar\ with $\r0 \gtrsim$ 1~km, we adopt representative values for other 
variables and set $e$ = 0.01:
\begin{eqnarray*}
M_{d, eq} & \approx & 7 \times 10^{18}~{\rm g} 
\left ( { \dot{M} \over 10^{10}~{\rm g~s^{-1}} } \right )^{1/2}
\left ( { 0.6~\msun \over \mwd } \right )^{9/20}
\left ( { \r0 \over {\rm 1~km} } \right )^{1.04}
\left ( { \rho \over {\rm 3~g~cm^{-3}} } \right )^{9/10} \\
 & & ~~~~~~~~~~~~~~~~\left ( { 0.01 \over e } \right )^{4/5}
\left ( { \Delta a } \over {0.2 a} \right )^{1/2}
\left ( { a \over \rsun } \right )^{43/20} ~~~~~~~~~~ \r0 \gtrsim {\rm 1~km} ~ . ~~~~ (11)
\label{eq: meq}
\end{eqnarray*}
\addtocounter{equation}{1}
The equilibrium mass is sensitive to the location of the ring. At fixed $a$,
$M_{d,eq}$ varies roughly linearly with \r0\ and as the square root of $\dot{M}$.

Maintaining this equilibrium requires that the time scale to replenish the 
ring, $t_r = M_{d,eq} / \dot{M}$, is not much longer than the collision time. 
For large objects and small $\dot{M}$, the mass in solids required to begin
the cascade is large; $t_r$ is also large. In these situations, we expect 
the ring to grow slowly in mass until the cascade begins; collisions then
rapidly deplete the ring. The total mass oscillates.

As the evolution proceeds, the vertical scale height depends on the time 
scale for mass to flow from \rmax\ to \rmin. In most cascades, material 
flows from \rmax\ to \rmin\ in a few collision times.  In some systems, 
however, there is enough mass in small particles for collisional damping 
to reduce the vertical scale height substantially \citep{kb2015a,kb2016a,
kb2016b}. For rings of solids around white dwarfs, the most massive 
equilibrium rings require high input rates of massive particles. Thus, 
these systems have the best chance of developing the physical conditions 
that promote collisional damping.

To test these ideas, we consider the evolution of a ring of solids with 
initial mass \M0\ = 0. Particles with radius \r0, $e_0$ = 0.01, and 
$\imath_0 = e_0/2$ are added to the ring at a rate \mdotz. In each time
step of length $\Delta t$, the number of particles with mass \m0\ added 
to the grid is $\Delta N = \mdotz\ ~ \Delta t / \m0$. Our algorithm uses 
a random number generator to round $\Delta N$ up or down to the nearest 
integer.  For systems with large \r0, this procedure introduces some shot
noise into the input rate. 

Each calculation follows the same pattern. Large solids are added to the 
swarm until they reach a critical cross-sectional area and begin to collide. 
Debris produced from the first collision then interacts with all other 
particles in the grid.  As this swarm evolves, large objects are continually 
added to the grid at the nominal rate \mdotz. These new objects continue to 
power the cascade.

Systems with small \r0\ and large \mdotz\ easily attain an equilibrium where
the mass and cross-sectional area of the swarm are roughly constant. In these
calculations, there is little shot noise in the collision rate among the particles
in the swarm or in the input rate of large objects. The equilibrium (\md, \ad)
depend on (\r0, \mdotz). 

\begin{figure} 
\includegraphics[width=6.5in]{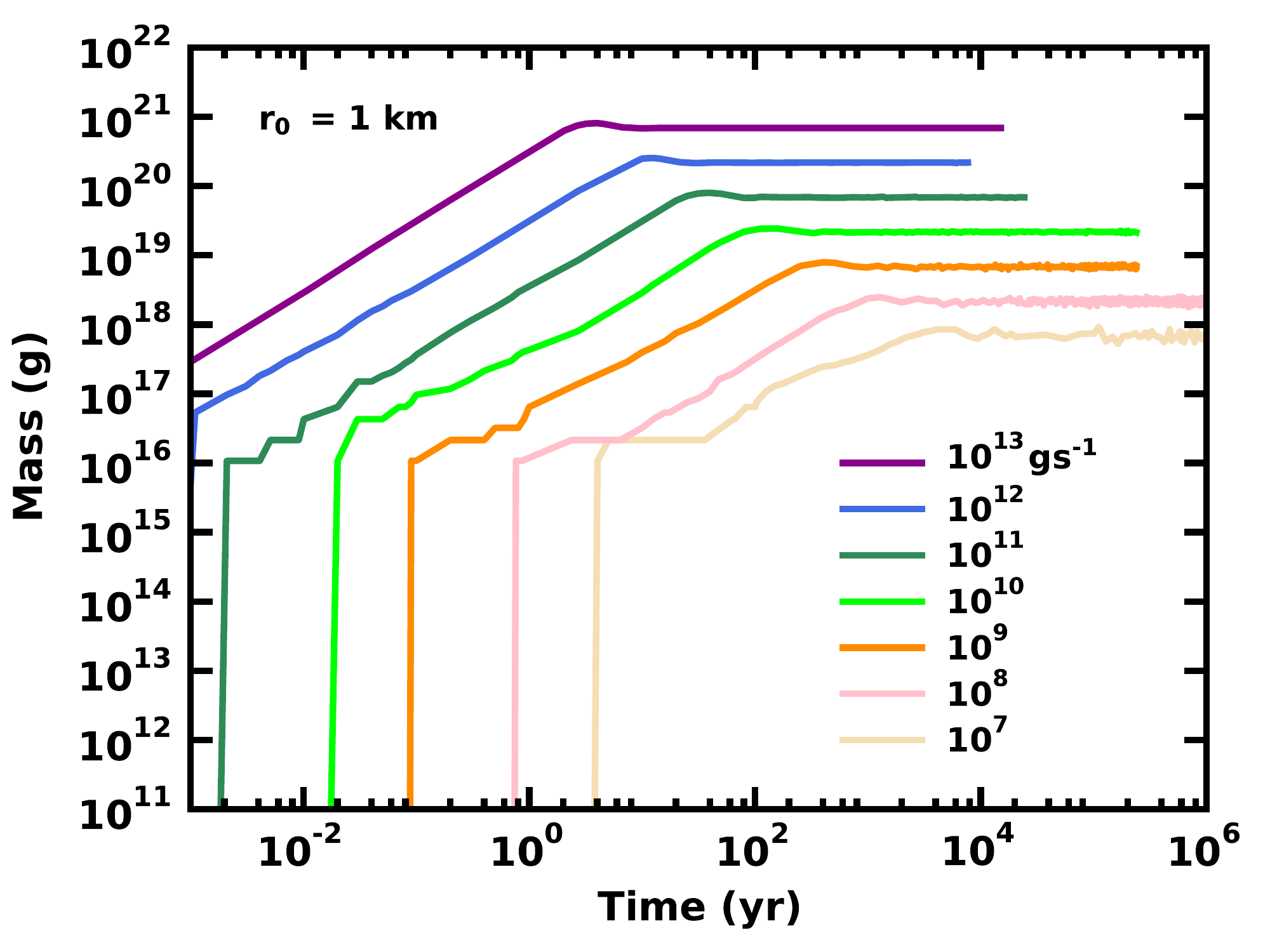}
\vskip 3ex
\caption{
Time evolution of the mass in solids for systems with zero initial mass 
where 1~km particles are added at rates indicated in the legend. 
All systems reach an equilibrium mass $M_{d, eq}$ which depends on the 
input rate \mdotz; fluctuations about $M_{d, eq}$ grow with decreasing 
\mdotz.
\label{fig: mass2}
}
\end{figure}

Fig.~\ref{fig: mass2} shows the evolution in mass for systems with \r0\ = 1~km
(\m0\ $\approx 10^{16}$~g).  Swarms always find an equilibrium where the mass 
is nearly constant in time. Oscillations about this equilibrium mass are negligible 
(modest) for large (small) input rates. Shot noise produces these oscillations. 
At the lowest (highest) rates, a new object is added every 30~yr (20 min).  At the 
lowest rates, a small degree of shot noise disrupts the smooth transport of mass 
from the largest to the smallest objects.

Systems with large \r0\ and small \mdotz\ cannot find an equilibrium. Shot
noise dominates the evolution.  The swarm repeats a standard sequence of
events, where (i) material is added until the cascade begins,
(ii) the cross-sectional area rises dramatically, (iii) a robust cascade
depletes the small particles in the swarm faster than collisions of large objects
can replenish them, and (iv) \ad\ and to a lesser extent \md\ decline dramatically.
The duty cycle of this process depends on \r0\ and the input rate. In these 
systems, the maximum mass is within a factor of ten of the equilibrium mass 
in eq.~\ref{eq: meq}.

\begin{figure} 
\includegraphics[width=6.5in]{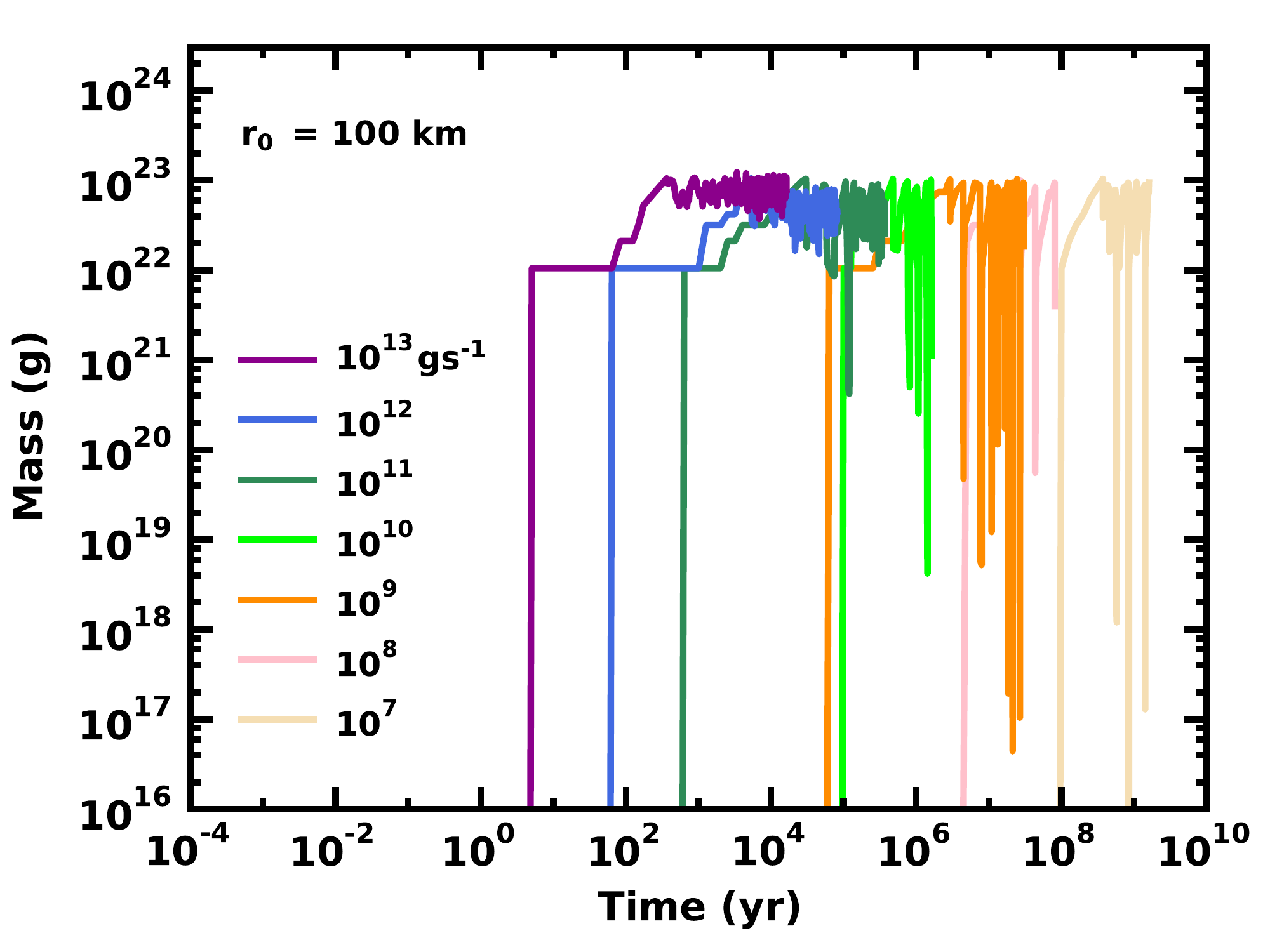}
\vskip 3ex
\caption{
As in Fig.~\ref{fig: mass2} for swarms with \rmax\ = 100~km. In swarms
with very large solids ($r \gtrsim$ 100~km), all systems with input 
$\dot{M} = 10^7 - 10^{13}$~\gs\ reach roughly the same maximum mass.
Fluctuations in the mass grow with decreasing input $\dot{M}$.
\label{fig: mass3}
}
\end{figure}

Fig.~\ref{fig: mass3} repeats Fig.~\ref{fig: mass2} for \r0\ = 100~km 
(\m0\ $\approx 10^{22}$~g). At the highest (lowest) input rates, an object is 
added every 30~yr (every 30~Myr). It takes 30--40 large objects to commence 
the cascade. The mass in solids is then sensitive to the input rate. At the 
largest rates, shot noise produces variations in the rate mass flows down the 
cascade. Thus, the equilibrium mass varies. At the lowest rates, the system 
gradually grows in mass until it has the requisite number of large objects 
to begin the cascade. Collisions then depletes the system at a rate much 
faster than the rate of adding large objects to the swarm. The mass drops and 
remains at some minimum level until the input of large objects raises the mass 
to the level required for a cascade.

When \mdotz\ is larger, the systems come closer to reaching an equilibrium.
Large drops in the total mass become more and more infrequent. Once the input
rate reaches $10^{12}$~\gs, the solids find a rough equilibrium with a mass
close to the expected mass from eq.~\ref{eq: meq}. Shot noise in the input 
rate generates fluctuations about the equilibrium. 

For systems with large \r0, all input \mdotz\ lead to roughly the same maximum
mass in solids. This mass is close to the equilibrium mass for high input \mdotz.
In these rings, the mass in solids grows until the collision rate among the high 
mass objects reaches the rate required to power the cascade. This rate only depends
on the cross-sectional area of the largest objects. While the rate is sensitive
to \r0, it is independent of \mdotz. 

When \r0\ = 1--100~km, the equilibrium disk mass agrees amazingly well with
the analytical prediction (eq.~\ref{eq: meq}). To make this comparison we 
examine the mass ratio $\xi = M_{eq, n} / M_{d, eq}$, where $M_{eq, n}$ is 
derived from simulations and $M_{d, eq}$ is the analytical prediction 
(eq.~\ref{eq: meq}). For large \r0, we consider only those calculations with
large \mdotz\ where the fluctuations in the mass are small. Among 46 simulations
with \mdotz\ = $10^7 - 10^{13}$~\gs, $\xi$ = 2.0--2.5 (Fig.~\ref{fig: meq});
the average ratio is $\bar{\xi} = 2.19 \pm 0.15$.  Despite the factor of two 
offset, the numerical simulations match the predicted variation of the
equilibrium mass with \mdotz\ and \r0.  Considering the approximations made in 
deriving the analytical prediction, the good agreement is more than satisfactory.

Within a large suite of simulations with \r0\ = 0.1--300~km and 
\mdotz\ = $10^7 - 10^{13}$~\gs, collisional damping has negligible
impact on the evolution. Fig.~\ref{fig: damp3} shows an example of
the evolution for \r0\ = 10~km and \mdotz\ = $10^{13}$~\gs.  When 
the cascade begins, the solids have a vertical scale height $H$ = 2800~km. 
As the simulation proceeds, large objects with $r \gtrsim$ 0.1--1~km 
maintain this scale height. Collisional damping is ineffective.
Although collisional damping often reduces the vertical scale height 
of small objects with $r \lesssim$ 1--10~m, the reduction is modest. 
Once systems reach the equilibrium mass, small particles have 
$H \approx$ 1250--1750~km. When systems cannot reach an equilibrium, 
collisional damping is more sporadic; $H \approx$ 2500--2800~km.

In all simulations, the reprocessed luminosity of the particles 
\ld\ closely follows the total mass \md. In Fig.~\ref{fig: lum2}, the
predicted \ldlwd\ for systems with \r0\ = 1~km rises when the solid mass 
reaches $\md \approx 10^{18}$~g. Shortly thereafter, the luminosity finds 
an equilibrium with 
\ldlwd\ $\approx$ $10^{-3} (\dot{M}_0 / 10^{12}~{\rm g~s^{-1}})^{1/2}$. 
Systems with smaller \mdotz\ display larger fluctuations about this equilibrium
luminosity. 

Once \r0\ $\gtrsim$ 100~km, rings of solids with input \mdotz\ $\lesssim$ 
$10^{10}$~\gs\ spend most of their time in a very low luminosity state with
\ldlwd\ $\approx 10^{-8} - 10^{-7}$ (Fig.~\ref{fig: lum3}). During this
period, the mass in large objects slowly grows. Eventually, the mass reaches
a critical level of roughly $10^{23}$~g. Collisions then rapidly generate a 
luminous system, which quickly fades back to the faint minimum. 

Rings with \r0\ = 100~km and larger \mdotz\ attain a stable state where the
luminosity fluctuates around a rough equilibrium. For any \mdotz, the typical 
luminosity is identical to the \ldlwd\ achieved by systems with smaller \r0. 
However, the large flares in \ldlwd\ are much larger than those in rings with smaller \r0.

To quantify how often these calculations generate detectable IR excesses, 
we define the detection probability as the fraction of time where the 
reprocessed luminosity \ldlwd\ $\gtrsim 10^{-3}$.  Fig.~\ref{fig: ldet} 
summarizes our results. At large \mdotz, systems with \r0\ = 1--100~km 
spend nearly all of their time above the nominal detection limit. When 
\r0\ = 200--300~km, the time required to accumulate enough mass for the 
cascade is a significant fraction of the evolution time. Thus, these 
systems are rarely detectable even when \mdotz\ is large.

For intermediate accretion rates, \mdotz\ $\approx 10^{11} - 10^{12}$~\gs,
the detection probability is a few per cent. These systems spend more than
90\% of their time in low states, where the cascade is fairly dormant. Once
the cascade has enough mass, it briefly produces a detectable debris disk.

When the input \mdotz\ is small, the IR excess is rarely detectable. Cascades
with \r0\ $\lesssim$ 10~km never generate enough mass in small particles to 
reach \ldlwd\ = $10^{-3}$. Although systems with larger \r0\ sometimes achieve 
large \ldlwd, the fraction of time spent in the bright state is small.

These results are remarkably independent of other input parameters in our
calculations. The exponents $b_d$ and $b_l$ in our algorithms for debris
production have a limited impact on \md\ and \ld.  Adopted values for 
\qdstar\ are more important: increasing (reducing) \qdstar\ slows down 
(speeds up) the conversion of particles with $r \lesssim$ 1~km into 
smaller particles \citep{koba2010a,wyatt2011,kb2016a,kb2017}. Thus,
systems with smaller (larger) \qdstar\ have smaller (larger) equilibrium 
values for \md\ and \ld. For factor of three changes in \qdstar, however,
collisional damping still has negligible impact on the evolution. Cascades 
of destructive collisions of objects with \r0\ = 0.1--300~km {\it always} 
prevent small particles from attaining a small vertical scale height.

\section{DISCUSSION}
\label{sec: disc}

Our calculations are the first to quantify the collisional evolution of 
massive rings of solid particles near the Roche limit of a white dwarf. 
Including accurate treatments for collision outcomes, dynamical interactions
among the solids, and Poynting-Robertson drag, we derive the behavior of 
swarms as a function of various properties of the solids.

If ensembles of solids have no input from material outside the Roche limit, 
evolution is very rapid.  In $\lesssim 10^3$~yr, collisions transform systems 
of 1--100~km objects with initial surface density $\Sigma_0$ = 
100~\gcms\ (mass $M_0 \approx 10^{24}$~g) and orbital eccentricity $e_0$ = 0.01 
into 1~\mum\ particles which are swiftly vaporized into a metal-rich gas. 
Dynamical processes such as collisional damping, dynamical friction, 
Poynting-Robertson drag, and viscous stirring play a minor role throughout 
the evolution. 

Although choosing different initial masses for these systems changes the 
evolution time, long-term outcomes are identical.  Eventually, cascades 
with \r0\ = 1-100~km and $M_0 \lesssim 10^{25}$~g converge on the same 
size and velocity distributions. Dynamical processes remain unimportant.

Adding material to the ring at a constant rate \mdotz\ enables the system to 
reach an equilibrium where the mass of the swarm scales with $\dot{M}_0^{1/2}$ 
and \r0. The equilibrium mass derived from numerical simulations is close to 
the predictions of a simple analytical model. For \r0\ $\lesssim$ 10~km, these 
equilibria are stable.  Despite the roughly constant masses of these rings, 
dynamical processes still act on time scales much longer than the collision 
time. Thus, the vertical scale height $H$ remains larger than \r0.  In systems 
with larger \r0, the mass fluctuates by 1--10 orders of magnitude on time scales, 
$10^2 - 10^7$~yr, which depend on \mdotz. Dynamical processes are still relatively 
unimportant.

When rings of solids have an equilibrium mass and \mdotz\ $\gtrsim 10^{12}$~\gs,
the stellar luminosity reprocessed by small particles is comparable to the 
luminosity of IR excess emission observed around many metallic line white dwarfs
(Figs.~\ref{fig: lum2}--\ref{fig: lum3}). 
If \mdotz\ is smaller, adding large objects with $\r0\ \gtrsim$ 30~km to the 
swarm results in sporadic periods of large \ldlwd. The fraction of time spent
with \ldlwd\ $\gtrsim 10^{-3}$ ranges from much less than 1\% (\r0\ $\gtrsim$ 
30~km, \mdotz\ $\lesssim 10^8$~\gs) to 10\%--25\% 
(\r0\ $\gtrsim$ 30~km, \mdotz\ $\approx 10^9 - 10^{11}$~\gs).

\subsection{Collisional Damping in a Collisional Cascade}
\label{sec: disc-damp}

At first glance, the differences between Fig.~\ref{fig: damp1}
and Fig.~\ref{fig: damp3} might seem remarkable. In systems with 
indestructible particles, damping reduces the vertical scale height 
by 50\% in roughly ten collision times, 
$t_{50} \approx 10 t_0 \approx 10^4$~s.  
Once collisions become destructive, damping is negligible.  

However, destructible particles with $e_0 \approx$ 0.01 do not 
survive long enough to damp.  
When two 1~cm particles with $e \approx$ 0.01 collide, their center-of-mass 
collision energy is $Q_c \approx v^2/8 \approx 10^{10}$~\ergg.  With 
$ \qdstar = 3 \times 10^7$~\ergg, $Q_c / \qdstar \approx$ 375; the particles
are completely destroyed. From eq.~\ref{eq: mlarge}, the radius of the largest 
particle in the debris is roughly 1~mm. The collision of two 1~mm particles 
yields material with sizes of 0.13~mm and smaller. It then takes another 
3--4 collisions to reduce the debris to sub-micron sizes. Overall, the 
5--6 collisions required to convert a pair of cm-sized particles into
copious amounts of micron-sized particles is a factor of two smaller than
the number of collisions required to damp their velocities by 50\%. 

Collisions can reduce the vertical scale height of large objects with 
mass $m_1$ after interactions with an equivalent mass in smaller objects
\citep[e.g.,][]{gold2004}. When these large objects orbit with $e \approx$ 0.01 
at the Roche limit of a white dwarf, however, they lose mass faster than they 
damp. For example, a 10~km (100~km) object colliding with a single smaller 
particle of mass $m_2$ loses a mass $\Delta m_1 \approx 10^3 m_2$ 
($\Delta m_1 \approx 50 m_2$). Defining $N$ = $m_1 / m_2$ as the number of 
collisions required for the larger object to interact with an equivalent mass
in smaller objects, it is clear that the total mass lost ($N \Delta m_1$) is
much larger than $m_1$. 

Overall, the level of damping in Fig.~\ref{fig: damp3} agrees with expectations 
based on the survival times. Over 5--6 collisions among small particles, we 
anticipate a 25\% to 30\% reduction in the scale height, which is reasonably close
to the 33\% reduction derived in the calculations. Because collisions with small 
particles rapidly destroy larger particles, damping of 1~m and larger particles 
should be smaller than that of smaller particles, as shown in Fig.~\ref{fig: damp3}.
Once the mass in the annulus reaches a steady-state, the vertical scale height 
also reaches a steady-state which is set by the amount of damping achieved during
the fairly short residence times of the small particles in the grid.

Compared to other situations where damping {\it has} been effective during a
cascade \citep[e.g.,][]{kb2009,kb2015a,kb2016a,kb2016b}, the Roche limit of 
a white dwarf is very harsh. At 1--100~AU, $e \approx$ 0.01 produces modest 
ratios $Q_c / \qdstar \approx 1-3$ instead of the $Q_c / \qdstar \approx$ 
100--1000 discussed here. When $Q_c/\qdstar \approx 1$, it takes $\gtrsim$ 
25--50 collisions to reduce a pair of 1~cm particles into sub-micron debris.  
If it takes 10 collisions to reduce the vertical scale height by a factor of 
two, there is a reasonable chance that damping can reduce the vertical scale 
height before particles are ground to dust.  At the Roche limit, however, 
large and small particles are ground to sub-micron sizes much more rapidly 
than at 1--100~AU. Thus, damping is ineffective.

Since collisions and gravitational dynamics are unable to reduce $H$ significantly, 
it is important to consider other physical processes capable of circularizing
particle orbits. Within a protoplanetary disk, gas drag is a vital component of 
the growth of planetesimals into protoplanets \citep[e.g.,][]{youdin2013}.  The 
Yarkovsky and YORP effects modify the orbits of asteroids and satellites in the
solar system \citep[e.g.,][]{bottke2006}. In the next two sub-sections, we examine
whether any of these processes can circularize the orbits of particles involved 
in a cascade.

\subsection{Reducing the Vertical Scale Height: Gas Drag}
\label{sec: disc-drag}

To estimate the impact of gas, we consider a simple model for a steady gaseous 
disk fed at a constant rate by the vaporization of 1~\mum\ or smaller particles.
The disk surface density
$\Sigma_g = \mdotz/ 3 \pi \nu$ where $\nu = \alpha c_s H_g$ is the viscosity,
$\alpha$ is the viscosity parameter, $c_s$ is the sound speed, and $H_g$ is
the vertical scale height of the gas. Adopting $\alpha \approx 10^{-3}$ 
\citep{metzger2012} and a gas temperature $T_g$ = 4000~K at 
$a \approx$ 1~\rsun\ \citep[e.g.,][]{melis2010b,melis2012a,metzger2012}, 
$c_s \approx$ 1~\kms, $H_g \approx$ 2000~km, and $\Sigma_g \approx$ 
0.05~\gcms\ for \mdotz\ = $10^{10}$~\gs. This estimate is similar to 
the $\Sigma_g \lesssim$ 0.01--0.1~\gcms\ derived in numerical simulations 
\citep[e.g.,][]{rafikov2011b,metzger2012}.

In this example, the sound speed is comparable to the vertical velocity of
solid particles. With $H_g \approx H_s$, solids spend most of their time 
interacting with the gas. For simplicity, we assume the solids `see' a 
typical gas density, 
$\rho_g \approx \Sigma_g / H$ $\approx 3 \times 10^{-10}~\gcmc$. Factor 
of 2--3 changes in $\rho_g$ have little impact on our discussion.

Within the gas, drag forces circularize the orbits of small particles
\citep{ada1976,weiden1977a,raf2004,chiang2010,youdin2013}.  Defining $F_d$ 
as the drag force, the `stopping time' is $t_s = m v_g / F_d$, where $v_g$ is 
the velocity of a particle relative to the gas. For small particles, we consider
drag in the Epstein and Stokes regimes, which balance when the particle size
$r \approx 9 \lambda / 4$, where $\lambda = \mu m_H c_s P / 2 \pi \Sigma_g \sigma_c$ 
is the mean free path, $\mu = 28$ is the mean molecular weight, and
$\sigma_c = 5 \times 10^{-15}$~cm$^2$ is the collision cross-section for two 
Si atoms \citep{weiden1977a,raf2004,metzger2012}.  In our model disk, 
$\lambda \approx$ 4~cm; drag is in the Epstein (Stokes) regime for 
$r \lesssim$ 4~cm ($r \gtrsim$ 4~cm). At 10~cm, the relevant stopping time is
$t_s \approx 10^6$~s. The stopping time is comparable to the orbital period 
when $r \approx$ 1~mm. 

Knowledge of the stopping time allows us to assess the response of particles 
to the gas. When \mdotz\ $\approx 10^{10}$~\gs, particles with $r \approx$ 1~mm 
encounter a mass in gas roughly equal to their own mass every orbital period. 
These and smaller `coupled' particles become entrained in the gas and maintain 
a large vertical scale height. With velocities comparable to the sound speed, 
collisions between these particles are destructive and generate debris which 
remains coupled to the gas. Thus, the gas does not help to halt the cascade and 
reduce the vertical scale height of the smallest particles.

Although entrained particles drift radially, the drift time is longer than 
the collision time. For $\alpha = 10^{-3}$, the viscous time scale is roughly 
$10^3$~yr. In our calculations, $t_c \lesssim$ 1--10~yr. Thus, small particles 
undergo destructive collisions before they drift out of the ring.

Loosely coupled particles larger than 1~mm (i) drift radially inward and 
(ii) damp in $e$ and $i$. The gas has a finite pressure and orbits the 
central star at a lower velocity than solids on Keplerian orbits 
\citep{ada1976,weiden1977a}. Solids feel a `headwind' which drags them 
inward and circularizes their orbits.  Large particles with $r \gtrsim$ 
10~m feel little headwind and experience little radial drift or damping. 
The maximum radial drift rate for cm-sized to m-sized particles is the 
difference in orbital velocity between the gas and the solids. For 
metal-rich gas orbiting a white dwarf, this velocity is roughly 
1~\ms\ \citep{metzger2012}. The time scale to drift out of an annulus 
with width $\Delta a \approx 0.2 a$ is several years. Orbits circularize 
on a similar time scale.

To estimate the relevance of radial drift and circularization, we compare 
the drift time scale to the collision time (eq.~\ref{eq: t0}) and the 
time scale for the cascade to process small particles from 1--100~cm to 
1~\mum\ ($t_p \approx M_s / \mdotz$, where $M_s$ is the mass in small 
particles). For cascades with \r0\ = 1~km and \mdotz\ = $10^{10}$~\gs, 
$t_0 \approx$ $10^5$~s and $t_p \lesssim 10^6$~s. The drift time scale 
is much longer than the collision time and somewhat longer than the 
processing time.  Thus, it seems unlikely that gas drag has much impact 
on the cascade: collisions process small particles faster than the gas 
drags them inward. 

For swarms with the equilibrium mass, $M_{d,eq}$, changing \mdotz\ is 
unlikely to modify these conclusions. Although the radial drift time 
is insensitive to the gas density, gaseous disks with larger \mdotz\ damp 
small particles more rapidly, $t_{damp} \propto \dot{M}_0^{-1}$.  The 
collision and processing times scale as $\dot{M}_0^{-1/2}$. Systems with
\mdotz\ $\lesssim 10^{10}$~\gs\ are {\it less} susceptible to gas drag 
than those with larger \mdotz.  When \mdotz\ $\approx 10^{14}$~\gs, 
$t_{damp} \approx t_c$. Although the gas might then circularize the orbits 
of small particles before collisions can destroy them, this \mdotz\ is 
much larger than the upper end of the range of observed \mdot\ in metallic 
line white dwarfs. 

When \r0\ $\gtrsim$ 30~km, the maximum mass is comparable to the equilibrium 
mass for a large \mdotz\ (Fig.~\ref{fig: mass3}). The ability of gas to reduce 
the vertical scale height of small particles then depends on how the gas 
and the solids interact during the short period of time when the cascade 
operates. Although evaluating the outcome requires a time-dependent 
calculation of the gas and the solids, our expectation is that the gas cannot 
reduce the vertical scale height of 1--100~cm particles before collisions
them into mm and smaller particles which will become entrained in the gas 
and maintain a large vertical scale height.

Although improving our assessment requires more rigorous calculations of the 
gas, these results suggest that the gas has modest impact on the cascade. 
Combined with our earlier results, collisional dynamics, gas dynamics, and 
PR drag are unable to reduce the vertical scale height of solids from several 
thousand km to several km or several m.

\subsection{Reducing the Vertical Scale Height: Other Options}
\label{sec: disc-yark}

Aside from radiation pressure and PR drag, other interactions between stellar 
radiation and small solids generate changes in $a$ and $e$. In the Yarkovsky
effect, differences in the loss of radiation from the hotter parts of a 
rotating solid relative to the cooler parts change $a$ and to a lesser extent 
$e$ \citep[see the discussions in][and references therein]{burns1979,bottke2006}. 
Differential radiation loss from the irregular surface of a solid \citep[the 
YORP effect;][and references therein]{rubincam2000} modifies the spin rates of
small solids and hence alters $a$ and $e$. Although these effects change $a$
and $e$ on long time scales \citep{burns1979,bottke2006,veras2014b}, they 
have negligible impact over the typical collision or processing time. 

By analogy with the Yarkovsky and YORP effects, differential mass loss across 
the surface of a solid from sublimation might also induce $\dot{a}$ and 
$\dot{e}$ \citep{veras2015c}. Rapid sublimation of ices from solids modify 
$a$ and $e$ on hundred year time scales. For systems dominated by collisions, 
however, all ices are probably liberated from the solids in a few collisions.
Subsequent sublimation of the rocky material left behind occurs over much
longer time scales. 

Overall, radiation dynamics seems unlikely to reduce the vertical scale height 
of small solids orbiting at the Roche limit.

\subsection{Shrinking Collisionless Rings}
\label{sec: disc-shrink}

If these processes are unable to modify $e$ and $i$ significantly, it is 
necessary to identify other mechanisms.  As one possibility, \citet{veras2014a}
consider tidal disruption of a spherical non-rotating asteroid passing within
the Roche limit of the white dwarf \citep[see also][]{richard1998,hahn1998,
scheeres2000,sharma2006,movshovitz2012}.  The asteroid has a rubble-pile 
structure with modest or negligible internal cohesion 
\citep[e.g.,][]{richard2005,sharma2009}.  With these assumptions, asteroid 
disruption produces a set of distinct particles distributed along the original 
orbit of the asteroid.  The system is then an eccentric, collisionless ring 
of small solid particles.

Once the ring forms, radiation from the central white dwarf shapes the dimensions
of the ring \citep{veras2015b}. For 1~cm particles with $e_0$ = 0.99 at $a_0$ = 
10~AU, the time scale to drift inside 1~\rsun\ is roughly
600~$(r / {\rm 1~cm})~(L_{wd} / 10^{-2} \lsun)^{-1}$~Myr \citep{burns1979,veras2015b}.  
As particles pass inside the Roche limit, they have $e_R \approx$ 0.01.  These 
outcomes are very sensitive to $e_0$. When $e_0$ = 0.999 (0.98), the drift time 
drops (rises) to 20~Myr (1600~Myr) with $e_R \approx$ 0.16 (0.004). For a typical 
white dwarf luminosity, 1~cm particles that reach the Roche limit on a reasonable 
time scale have fairly eccentric orbits ($e \gtrsim$ 0.01).

In these circumstances, collisional evolution of ring particles seems inevitable.
At large $a$, particles on very eccentric orbits have shorter time scales for PR 
drag than for collisions. As $a$ and $e$ drop, $t_0$ gradually becomes shorter 
than $t_{PR}$ (see eqs.~\ref{eq: t0}--\ref{eq: tpr}).  Although it is possible that 
ring particles maintain a collisionless structure inside the Roche limit, small 
differences in the properties of ring particles probably generate a modest 
dispersion in $a$, $e$, and $\imath$ as the orbit contracts from $a \approx$ 
10~AU to $a \approx$ 1~\rsun. This dispersion leads to crossing orbits and 
collisional evolution.


As an example, tidally disrupting a single 1~km asteroid yields roughly $10^{16}$ 
1~cm particles with a total cross-sectional area $A_d \approx 10^{16}$~cm$^2$. 
For orbits with $e \approx$ 0.01 (0.1), $a \approx$ 1~\rsun, and
$\Delta a \approx e a$, the collision time is a few months (a few yr; eq.~\ref{eq: t0}), 
which is much shorter than the many centuries required for radiation to pull 
the solids into the sublimation radius. Because the collision time and the drag 
time both scale with particle radius, collisions always dominate.

From our calculations, rings of small particles with $e \gtrsim 10^{-3}$ produce
a cascade which grinds solids into smaller and smaller objects.  Although this 
evolution cannot modify $e$ significantly, collisions try to puff up the 
inclinations of ring particles until $i \approx e/2$. The time scale for this 
change is 5--10 collision times, roughly the time to grind cm-sized to m-sized
objects to dust.

\subsection{Evolution of Solids: Alternative Treatments}
\label{sec: disc-other}

To develop a better understanding of the relationship between solid particles 
and gas in white dwarf debris disks, Rafikov and collaborators investigated
models where solids interact with the radiation field of the white dwarf and a 
gaseous circumstellar disk \citep{rafikov2011a,rafikov2011b,bochkarev2011,metzger2012,
rafikov2012}. In their picture, ensembles of small solids with $r \sim$ 1~cm 
produced by tidal disruption of a single, much larger object follow circular 
orbits within a disk or ring inside the Roche limit.  PR drag pulls these 
particles closer to the white dwarf, where the radiation field eventually 
vaporizes them. Vaporization generates a gaseous disk, which then interacts 
with the small particles through gas drag. On time scales somewhat longer 
than the viscous time of $10^3$ yr, the system may develop phases of runaway 
accretion, where the rate gas falls onto the white dwarf grows dramatically 
with time. The runaway depletes the gaseous disk. If enough solids remain,
this process can repeat; otherwise, a new runaway requires disruption of 
another large object.

In this approach, interactions among the small solids are minimal 
\citep[see also][]{farihi2008}. For cm-sized objects on circular orbits, 
collisional damping maintains low $e$ and low $\imath$. Any low velocity 
collisions probably produce rebounds with little or no debris; mergers are 
rare. Although repeated rebounds gradually spread the ring, this process 
is slow. Under these conditions, there is little likelihood of a cascade 
or other rapid evolution of the solids.

From our calculations, attaining the initial configuration of this model 
seems challenging.  Small solid particles generated from a disrupted 
asteroid or comet probably have modest $e$ inside the Roche limit 
\citep[e.g.,][]{veras2015b}. Once the orbits of these particles cross, 
a cascade is inevitable. The solids are then rapidly ground to dust and 
vaporized into a gas. From our estimates, mm-sized to cm-sized solids 
interacting with the gas do not drift very far from the cascade. 
Thus, it is hard to produce a swarm of small solids on nearly circular 
orbits.

Understanding the orbital geometry of solids inside the Roche limit requires
more comprehensive calculations of the evolution of solids produced during
the disruption of a comet or asteroid. If these solids can damp onto circular
orbits before a cascade begins, then small particles can feed the structures
envisioned in \citet[][and references therein]{rafikov2012} and generate gas
which eventually accretes onto the central star. Otherwise, collisions are a
more likely source of a gaseous disk. 

\subsection{Contacts with Observations}
\label{sec: disc-obs}

Despite the general failure in providing a clear path to disks with a small 
vertical scale height, cascade calculations of solids at the Roche limit 
enrich our understanding of metallic line white dwarfs. Although a 
disintegrating asteroid is a popular mechanism \citep[e.g.,][and references 
therein]{jura2003,jura2008,bear2013,farihi2013,vanderburg2015,farihi2016,
gurri2017}, cascades initiated by collisions of large asteroids (as investigated 
in \S3) or by collisions of debris from a tidal disruption are a plausible 
alternative. Rather than attempt to explain observations, here we consider
points of contact between our predictions and existing data.  As we include 
more physical processes in the simulation, we plan more comprehensive 
comparisons with real systems.

\begin{itemize}

\item Dusty material with a large vertical extent is necessary to produce 
the broad range of eclipses in WD1145+017 \citep{vanderburg2015,gansicke2016,
alonso2016,rappaport2016,zhou2016,croll2017,gary2017,hallakoun2017}.  
In a cascade interpretation, 
eclipses result from small particles in the debris from high velocity 
collisions. Stochastic collisions lead to debris with variable optical depth, 
accounting for short time scale variations in the system. Sublimation of 
particles with $r \lesssim$ 1~\mum\ explains the presence of strong absorption
features from a circumstellar gaseous disk. Entrainment of small particles 
within sublimating gas might account for plumes of material surrounding large
objects.  We plan additional calculations to consider this picture in more 
detail.

\item In other systems, dust with a large vertical scale height can account 
for IR excess emission with reprocessed luminosity \ldlwd\ $\approx$ $10^{-3}$ 
to $3 -4 \times 10^{-2}$ \citep[e.g.,][and references therein]{farihi2016}. 
Cascades with high \mdotz\ or large \r0\ account for these systems. Because 
these systems must wait to collect enough material to begin the cascade, 
models with large \r0\ and a broad range of \mdotz\ naturally explain the 
low frequency of DBZ/DAZ/DZ white dwarfs with detectable IR excesses or
gaseous disks.  Our 
calculations predict stochastic changes in the IR excess with a broad range
of amplitudes and time scales. Although we considered geometries where solids 
are confined to a narrow annulus, solids with a larger range in $a$ probably
follow a similar evolution \citep[e.g.,][]{kb2008,kb2012,knb2016}.  

\item In the cascade picture, metallic line white dwarfs without IR excesses
are systems with \mdotz\ $\lesssim 10^{11}$~\gs\ or systems with \r0\ $\gtrsim$
30~km in a `low' state between periods of intense collisions. From our
perspective, placing limits on the frequency of white dwarfs with circumstellar
gas but no IR excess constrains models for sublimation of solids and viscous 
transport of the resulting gas.

\end{itemize}

\subsection{Future Prospects}
\label{sec: disc-future}

From our calculations, it seems possible to incorporate collisional cascades
into the current framework for the formation and evolution of debris orbiting
metallic line white dwarfs. As high eccentricity debris from disrupting 
asteroids or comets settles into lower $e$ orbits, destructive collisions 
may play a useful role in generating a detectable IR excess. Gas from the 
vaporization of micron-sized particles within the cascade provides a path 
to produce the gaseous disks observed in many systems.

Developing a more robust model for cascades within the Roche limit
requires addressing several issues.  Solids with a large vertical scale height 
are hotter and more susceptible to rapid sublimation than solids in a disk with 
negligible vertical scale height \citep[e.g.,][and references therein]{jura2003,
farihi2016}. As discussed earlier, the magnitude of this effect depends on the 
vaporization time $t_v$ relative to the collision time $t_c$ and the time scale 
to replenish the cascade $t_r = M_d / \dot{M}_0$.  In most of our calculations, 
$t_v \gg t_c$ and $t_v \ll t_r$.  A proper treatment of vaporization requires 
(i) identifying mechanisms to feed the cascade and
(ii) a robust calculation of the evolution of the gas along with the evolution 
of the solids. 

It is also important to consider outcomes for cascades outside the Roche 
limit.  In our experience, collisional damping is effective at 0.1--2~AU
\citep[e.g.,][]{kb2016a,bk2017}. Coupled with the larger equilibrium mass 
at larger $a$ ($M_{d, eq} \propto a^{43/20}$, eq.~\ref{eq: meq}) and the less 
restrictive conditions for growth by merger \citep{weiden1984a,ohtsuki1993,
canup1995,porco2007}, the cascade is probably less efficient and much less 
luminous at 2--3~$a_R$ than at the Roche limit.  Investigating outcomes 
for collisional evolution at 1--10~$a_R$ would enable better comparisons 
between observations and numerical calculations.

Including the evolution of the gas in cascade calculations is another goal.
Although our estimates suggest the gas has a modest impact on the cascade,
inward drift of small particles entrained in the gas might increase the 
predicted magnitude of the IR excess relative to calculations without the
gas.

We also need a better understanding of the orbits, shapes, sizes, and spin 
characteristics of solids scattered into orbits with periastra inside $a_R$.
In many applications, collisional and dynamical evolution erases the initial 
state of the system; outcomes often have little relation to the initial 
conditions \citep[e.g.,][]{kl1998,kb2004a,kb2008,kb2010}. In the calculations 
described here, the largest objects do not grow and dynamical evolution is
negligible. Thus, outcomes are more sensitive to the initial $e$ and $\imath$
of solid material. More robust predictions for the distribution of orbital 
parameters for solids would allow better tests of cascades models for white
dwarf debris disks.

\section{SUMMARY}
\label{sec: summary}

We consider the collisional evolution of ensembles of asteroids orbiting 
within a narrow ring at the Roche limit of a low mass white dwarf.  Solids 
with initial radius \r0\ = 1--100~km and small eccentricity, $e \approx$ 
0.01, undergo a series of destructive collisions which grinds them to dust. 
As in other cascades, the dust luminosity declines roughly linearly with 
time \citep[e.g.,][]{wyatt2002,dom2003,kb2004a,wyatt2008,kb2016a,kb2017}. 

\begin{figure}
\includegraphics[width=6.5in]{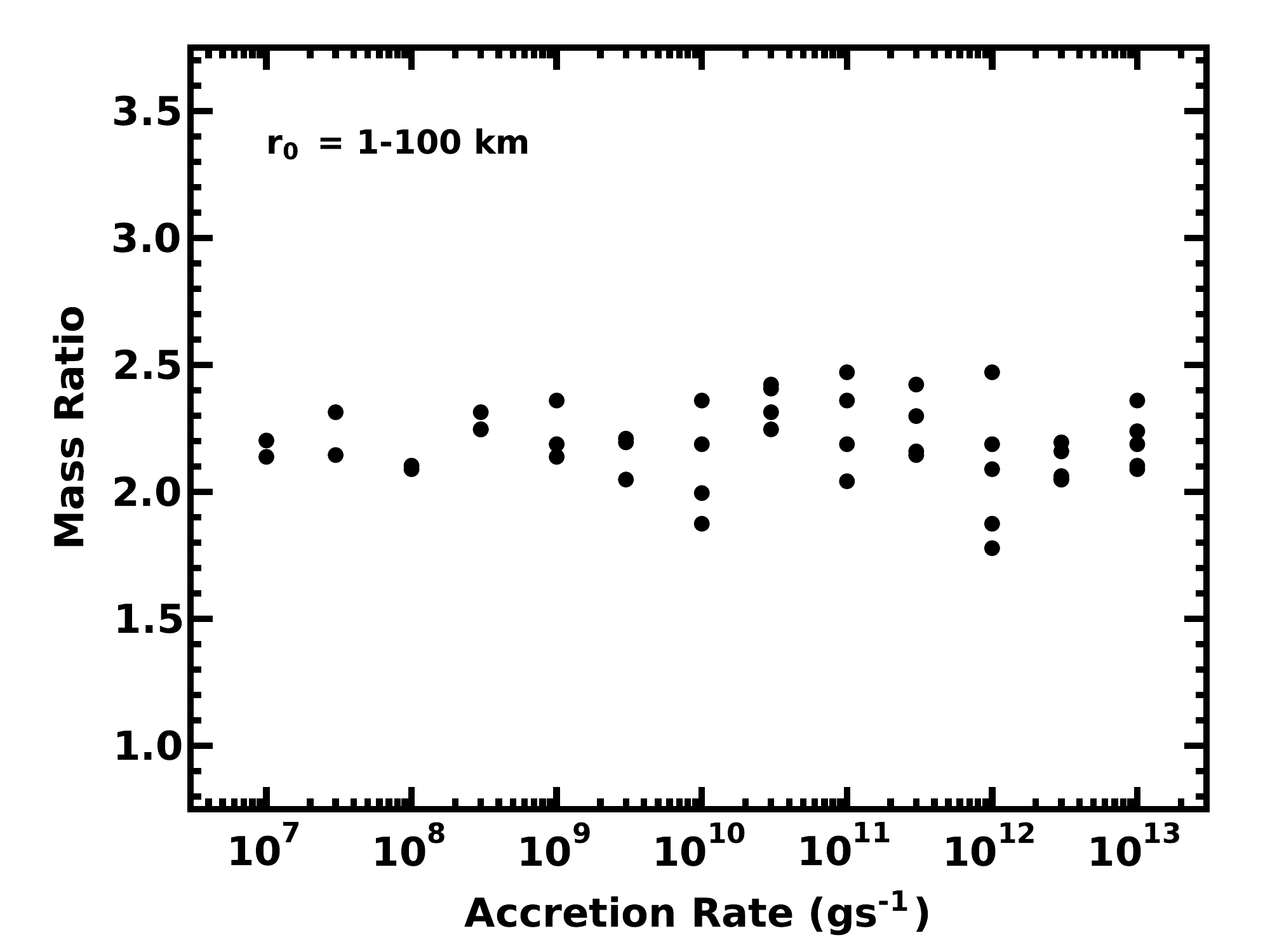}
\vskip 3ex
\caption{
Variation of the mass ratio $\xi = M_{eq, n} / M_{d, eq}$ as a function
of \mdotz\ for calculations with \r0\ = 1--100~km. Although the 
equilibrium mass derived in the numerical simulations is somewhat more 
than a factor of two larger than the analytical prediction, the 
calculations match the variation of the equilibrium mass with \mdotz\ and
\r0.
}
\label{fig: meq}
\end{figure}

\begin{figure}
\includegraphics[width=6.5in]{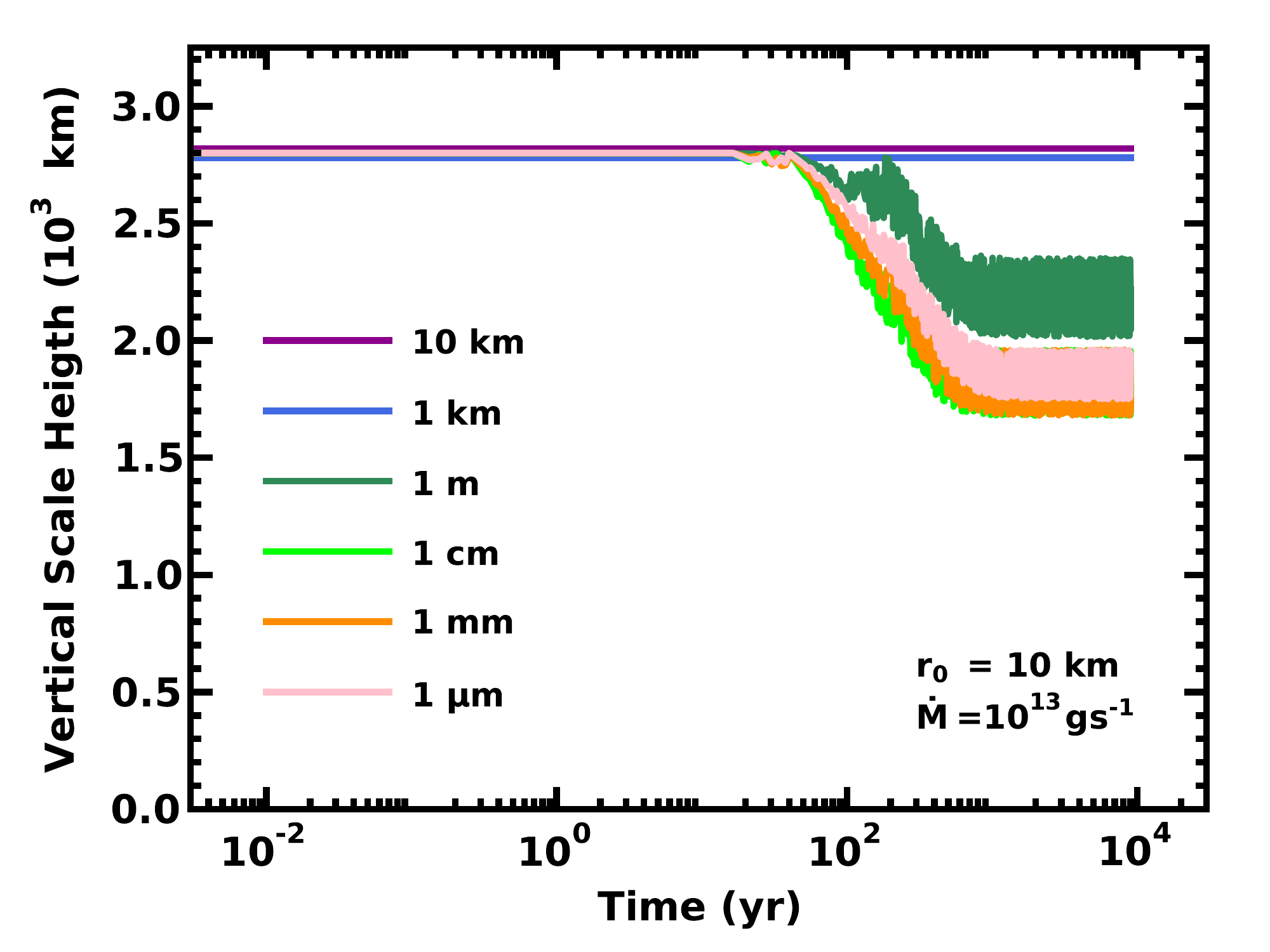}
\vskip 3ex
\caption{
Time variation of the vertical scale height as a function of particle size 
for a calculation with \r0\ = 10~km and \mdotz\ = $10^{13}$~\gs. Particle
sizes are indicated in the legend. As the mass in the annulus increases from
zero at $t$ = 0 to $7 \times 10^{21}$~g at $t$ = 40--50~yr, collisional 
damping is ineffective. Once the mass in solids reaches equilibrium, damping
gradually reduces the vertical scale height for particles with $r \lesssim$~10~m.
Damping is most effective for $r \approx$ 1--10~mm.
}
\label{fig: damp3}
\end{figure}

\begin{figure} 
\includegraphics[width=6.5in]{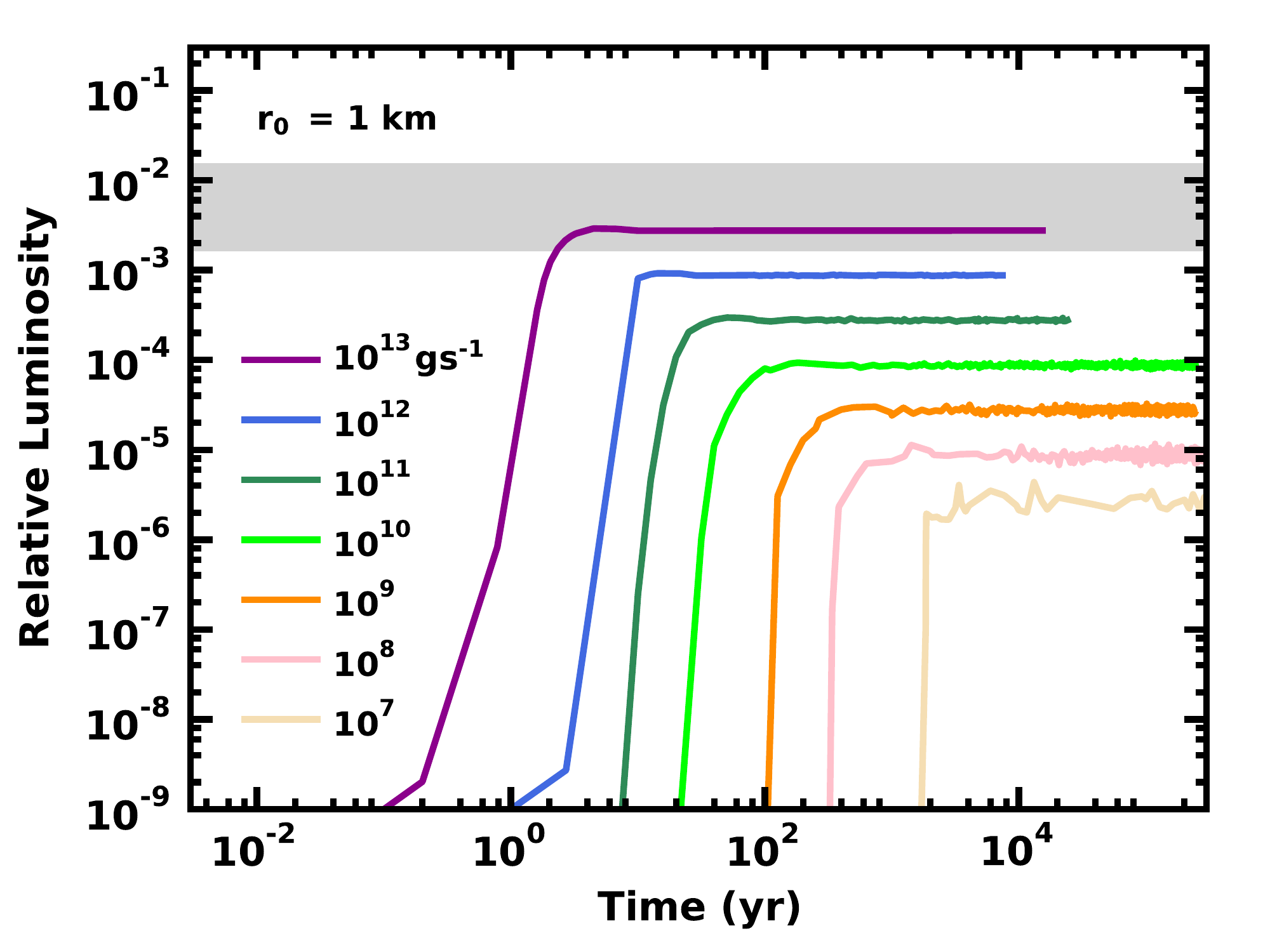}
\vskip 3ex
\caption{
Evolution of the reprocessed luminosity for swarms of solids with an 
initial mass of zero, \rmax\ = 1~km, and various input $\dot{M}$ as 
indicated in the legend. The grey bar indicates the observed range of
\ldlwd\ for white dwarfs with infrared excess emission. Although all 
systems reach a plateau luminosity which scales with $\dot{M}$, only
systems with $\dot{M} \gtrsim 10^{12}$~\gs\ achieve \ldlwd\ close to 
observed limits. Shot noise in the collision rates generate 
fluctuations about the plateau luminosity.  Systems with smaller 
$\dot{M}$ have larger fluctuations.
\label{fig: lum2}
}
\end{figure}

\begin{figure} 
\includegraphics[width=6.5in]{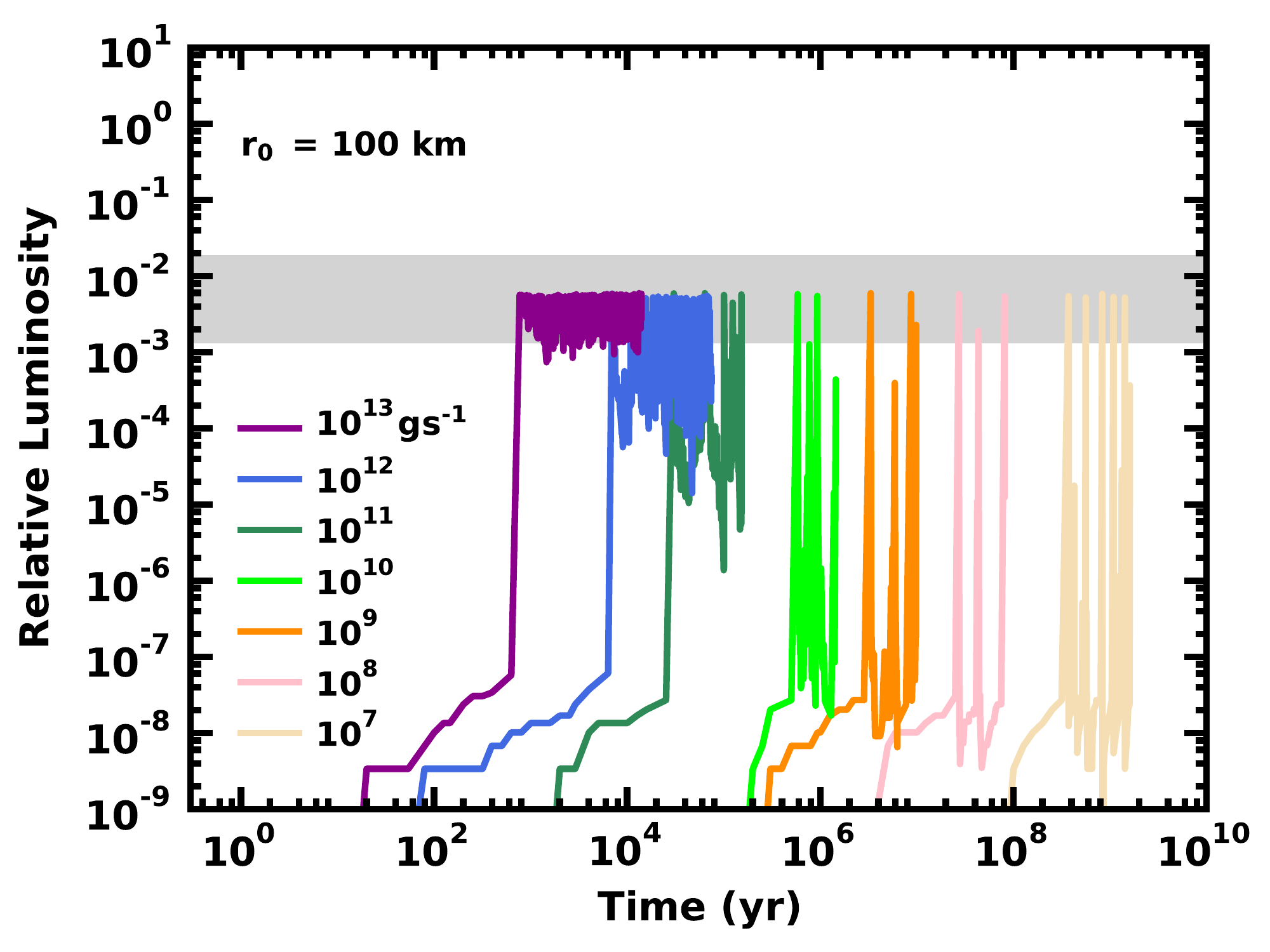}
\vskip 3ex
\caption{
As in Fig.~\ref{fig: lum2} for swarms with \rmax\ = 100~km.  For any
$\dot{M} = 10^7 - 10^{13}$~\gs, collision cascades occasionally 
generate enough small particles to match typical observed luminosities.
The fraction of time spent above \ldlwd\ $\gtrsim 10^{-3}$ scales
with the input $\dot{M}$.
\label{fig: lum3}
}
\end{figure}

\begin{figure} 
\includegraphics[width=6.5in]{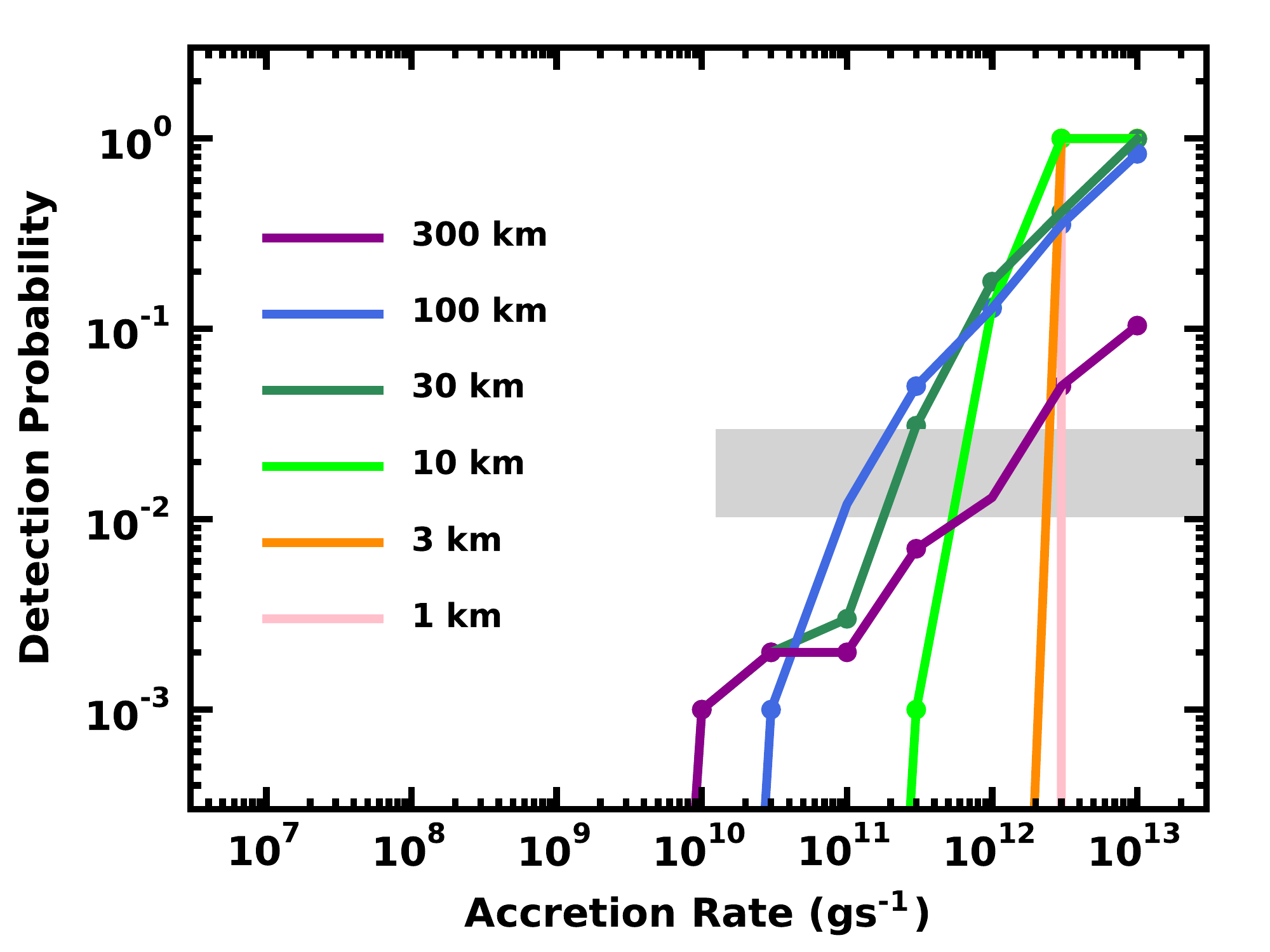}
\vskip 3ex
\caption{
Detection probability (fraction of time with \ldlwd\ $\gtrsim 10^{-3}$) 
as a function of \mdotz\ for various \r0\ listed in the legend.  The 
grey horizontal bar indicates the observed frequency of debris disks 
around metallic line white dwarfs. Input accretion rates 
\mdotz\ $\approx 10^{11} - 10^{12}$~\gs\ yield a detection probability
comparable with the observed rate.
}
\label{fig: ldet}
\end{figure}

When some physical process adds solids to the ring at a rate \mdotz, 
analytical results predict that the system evolves to an equilibrium mass 
which depends on \mdotz\ and the typical radius \r0\ of large objects 
added to the ring (eq. \ref{eq: meq}). For \r0\ $\lesssim$ 10~km, 
numerical calculations confirm this result (Fig.~\ref{fig: meq}). Adding 
larger objects to the ring leads to a system which
(i) maintains the equilibrium mass (for large \mdotz) or
(ii) oscillates between a very low mass and a high mass close 
to the equilibrium mass (for small \mdotz).

Solid material in rings with \r0\ = 1--100~km and total masses 
$M_d \approx 10^{20} - 10^{24}$~g evolve very rapidly. The typical 
time scale for collisions to convert sets of 1--100~km objects into 
1~\mum\ dust grains, $\lesssim$ 10--100~yr, is often much shorter than
the time scale for Poynting-Robertson drag to remove dust from the ring.
During this evolution, dynamical processes do not act fast enough to 
change $e$ and $\imath$ for solid particles. 

Throughout the cascade, other processes also seem incapable of changing 
$e$ and $\imath$. The Yarkovsky and YORP effects modify $e$ and $\imath$ 
on time scales much longer than the collision time.  In cascades that 
generate significant amounts of gas, small particles with $r \lesssim$ 
1~mm are probably well-mixed with the gas. The large vertical scale height 
of the gas guarantees that small particles maintain large $e$ and $\imath$ 
as they are ground to dust. The time scale for larger particles with 
$r \gtrsim$ 1~mm to drift radially inward is long compared to the 
collision time. Thus, gas has little impact on the cascade.

These results indicate that collisional, gas dynamical, and radiative 
processes cannot significantly reduce the vertical scale height of
swarms of solids with $H/R \gtrsim$ 0.01 before a cascade grinds solids 
into small dust particles. If disks of small solid particles orbiting 
white dwarfs have negligible vertical scale height, either the particles 
arrive with negligible $H$ or some other process circularizes their orbits.

Although it is premature to make detailed comparisons between the results 
of cascade calculations and data for DBZ/DAZ/DZ white dwarfs, it is encouraging 
that there are some points of contact between theoretical predictions and 
observations. The ability of cascades to maintain swarms of particles with 
large vertical scale height provides some hope for matching the IR excesses 
observed in many metallic line white dwarfs and the light curves of WD 1145+017.  
In cascades with large \r0\ and \mdotz, intermittent IR excess emission may 
improve insight into the origin of the relatively low frequency of metallic 
white dwarfs with IR excesses or gaseous circumstellar disks and the apparent 
stochastic nature of metal accretion onto DA and DB white dwarfs 
\citep[e.g.,][]{wyatt2014,bonsor2017}.

Developing a more predictive theory of collisional cascades at the Roche 
limit requires (i) expanding the single annulus model to a multi-annulus model 
spanning a broader range of semimajor axes,
(ii) including a prescription for evaporating/sublimating small solids into
gas and following the evolution of the gas and solids together, and 
(iii) constructing algorithms for feeding the coagulation code with results from 
dynamical simulations. Applying our multi-annulus treatment of coagulation and
gas dynamics to physical conditions near the Roche limit of a white dwarf is 
straightforward \citep[e.g.,][]{kb2008,kb2010,bk2011a,bk2013,kb2014}. Currently, 
there are many theories for bringing solids to the white dwarf 
\citep[e.g.,][]{debes2012a,frewen2014,stone2015,veras2015a,veras2015b,
veras2015c,payne2016,antoniadou2016,payne2017}. As these calculations mature, 
we plan to incorporate their approaches/results into our coagulation code.

We acknowledge a generous allotment of computer time on the NASA `discover' cluster. 
Portions of this project were supported by the {\it NASA Outer Planets Program} 
through grant NNX11AM37G. We thank S. Rappaport for suggesting we study white dwarfs 
with metallic absorption lines.  M. Payne provided a very useful set of comments
which helped us improve the manuscript.

\appendix

\clearpage

\bibliography{ms.bbl}

\clearpage


\clearpage


\clearpage

\clearpage

\clearpage

\clearpage


\clearpage

\clearpage

\clearpage


\clearpage

\end{document}